\documentclass{emulateapj}

\newcommand{\as}{\mbox{\arcsec}}

\newcommand\cmv{\mbox{cm$^{-3}$}}

\def\lsim {$\rlap{\raise.4ex\hbox{$<$}}\lower.55ex\hbox{$\sim$}\,$}

\newcommand{\msun}{\mbox{M$_\odot$}}

\input{epsf}

\def\c17o{$\rm C^{17}O$}
\def\dc18o{$\rm C^{18}O$}

\begin{document}

\title {\bf Evolution of Chemistry and Molecular Line Profile 
during Protostellar Collapse} 
\author {Jeong-Eun Lee}
\affil{\it Department of Astronomy, The University of Texas at Austin,
       1 University Station C1400, Austin, Texas 78712--0259}
\email{jelee@astro.as.utexas.edu}
\author {Edwin A. Bergin}
\affil{\it Department of Astronomy, The University of Michigan, 825 Dennison
Building Ann Arbor, Michigan 48109-1090}
\email{ebergin@umich.edu}
\and
\author {Neal J. Evans II}
\affil{\it Department of Astronomy, The University of Texas at Austin,
       1 University Station C1400, Austin, Texas 78712--0259}
\email{nje@astro.as.utexas.edu}

\begin{abstract}
Understanding the chemical evolution in star-forming cores 
is a necessary pre-condition to correctly assess physical conditions when
using molecular emission.
We follow the evolution of chemistry and molecular line profiles 
through the entire star formation process, including a self-consistent
treatment of dynamics, dust continuum radiative transfer, 
gas energetics, chemistry, molecular excitation,
and line radiative transfer. In particular, the chemical code follows
a gas parcel as it falls toward the center, passing through regimes of
density, dust temperature, and gas temperature that are changing both
because of the motion of the parcel and the evolving luminosity of the
central source.  We combine a sequence of Bonnor-Ebert 
spheres and the inside-out collapse model to describe dynamics from the
pre-protostellar stage to later stages. The overall structures of
abundance profiles show complex behavior that can be understood as 
interactions between freeze-out and evaporation of molecules. 
We find that the presence or absence of 
gas-phase CO has a tremendous effect on the less
abundant species.  In addition, the ambient radiation field and the grain
properties have important effects on the chemical evolution, and 
the variations in abundance have strong effects
on the predicted emission line profiles.  Multi-transition and multi-position observations are
necessary to constrain the parameters and interpret observations correctly
in terms of physical conditions. Good spatial and spectral resolution is 
also important in distinguishing evolutionary stages.
\end{abstract}

\section{INTRODUCTION}

In star-forming regions, the chemistry is not in steady state while cores are
being formed by contraction of parts of molecular clouds, or later, while the 
cores are collapsing.
Since star-forming cores can be studied through emission and absorption by
gaseous molecules, it is necessary to understand the chemical evolution
in order to correctly  assess the physical conditions.
Dust continuum emission is capable  of providing information 
on the core's physical structure and evolutionary state (Myers \& Ladd 1993, 
Andr\'e, Ward-Thompson, \& Barsony 1993), but velocity structure in the core, 
the other important physical parameter to constrain theoretical models in star 
formation, can only be obtained from molecular spectra (Mizuno et al. 1990; 
Zhou et al. 1993; Choi et al. 1995; Gregersen et al. 1997; 
Mardones et al. 1997; Gregersen et al.  1998; Di Francesco et al. 2001; 
Belloche et al. 2002). 
However, the observations of molecular spectra can be misinterpreted if
the molecular abundance distributions are not well understood.
For example, Rawings \& Yates (2001) caution that the blue asymmetry, 
which has been used to detect infall in star-forming cores, could disappear by 
the depletion of molecules in spite of infall velocity structure. 
Therefore, we need to understand the chemical evolution during star formation
to isolate good molecular tracers for each evolutionary stage and 
different regions along the line of sight. 

However, chemical processes in star-forming regions, especially reactions on 
grain surfaces and the gas-dust interaction, are poorly understood.
Several theoretical studies (Bergin \& Langer 1997; Aikawa et al. 2001; 
Caselli et al. 2002; Li et al. 2002; Shematovich et al. 2003; Aikawa et al. 
2003) examined the chemistry in pre-protostellar cores (PPCs). These
are considered as the potential sites of future star formation because they are
believed to be gravitationally bound, but have no central hydrostatic
protostar (Ward-Thompson, Scott, \& Andr\'e 1994; Ward-Thompson, Motte, \&
Andr\'e 1999). The only heating source of the PPCs is the interstellar radiation
field (ISRF), so the dust temperature in the inner denser regions is as low 
$\sim$7 K (Evans et al. 2001; Zucconi, Walmsley, \& Galli 2001). 
Those theoretical studies mentioned above modeled the gas-phase chemical 
evolution, including gas-dust interactions, and followed the changing 
depletions of some molecules and ionization fraction with time/density.
Each assumed that the gas temperature ($T_K$) is equal to the dust 
temperature ($T_D$) and constant through the cloud.
The consistent result from those studies is that sulfur- or carbon-bearing
molecules are easily depleted from the gas in dense inner regions during early stages, but 
nitrogen-bearing molecules become abundant at later times without significant
depletion.  
Based on the conclusion of those studies, we can consider nitrogen-bearing
molecules such as $\rm N_2H^+$ as good tracers of dense gas in evolved PPCs.
However, $\rm N_2H^+$ may also deplete (Bergin et al. 2002) 
in PPCs due to the long dynamical timescale compared to depletion timescale.
In addition, Lee et al. (2003) showed that the chemical evolution may depend on the 
dynamical scenario (e.g. free-fall or ambipolar diffusion) or 
different properties of dust grains as well as different density structures. 

In contrast to the situation in PPCs, there are few studies of
chemistry in cores after collapse begins. 
This is because the chemistry becomes more complicated due to additional 
heating mechanisms such as accretion luminosity or the luminosity from 
protostellar objects.
As a result of the additional heating sources, the dust temperature increases,
and icy materials will be evaporated from grain mantles, changing the available
molecular probes. 
Rawlings \& Yates (2001) tried to calculate the evolution of chemistry and line
profiles in Class 0 sources using a self-consistent chemical and dynamical 
model. They adopted the multi-point chemical model of Rawlings et al. (1992) 
in order to constrain the observed line profiles. 
However, they used only one dust temperature profile, 
obtained from dust continuum observations of B335 (Zhou et al. 1990),
assuming that the temperature profile does not vary with time.
In addition, they did not include gas energetics.  
Ceccarelli, Hollenbach, \& Tielens (1996) made the first attempt to calculate
self-consistently line spectra from the infalling gas at wavelengths between 
20 \micron\ to 200 \micron\ with a model that includes dynamics, chemistry, 
heating and cooling, and line radiative transfer.
In this work, they separated the dynamical and thermal evolution of the model
system. For dynamics, they adopted the simple solution of spherical isothermal
collapse (Shu 1977). The dust temperature was taken from the solution
of Adams \& Shu (1985) without calculating dust continuum radiative transfer.
In addition, they only considered OI fine-structure lines and CO and $\rm H_2O$
pure rotational lines as cooling mechanism, even though they included NIR
pump heating and dynamical compressional heating, which are important heating
sources at the inner hotter region, as well as gas-grain collisional heating.  
According to their results, the difference between dust and gas temperatures
is $\sim$25\% for the inner denser region, but the difference is larger 
in the outer layers. Other heating processes which could significantly 
contribute to the gas energetics, such as photoelectric heating and
cosmic-ray heating, were not considered by Ceccarelli et al.
For chemistry, they considered 182 pure gas-phase reactions including only 44 
species, most being oxygen- and/or carbon-bearing molecules. 
In their work, water vapor, at the inner region where $T_d \geq 100$ K, 
evaporated into the gas phase so that the cooling by $\rm H_2O$ rotational 
lines was enhanced significantly in the hot region. 
The most inconsistent assumption in this work with the results of studies for 
PPCs was the constant initial abundance profile of molecules.
They assumed the timescale of evolution before collapse was much shorter than 
the chemical timescale, so the chemical composition remained unchanged during 
the first phase of the condensation of the matter into a singular sphere. 
However, we know from observational studies and the theoretical models described
above that the abundance profiles in PPCs are not constant.

More recently, Rodgers \& Charley (2003) combined the dynamical evolution with
the chemical evolution focusing on the hot core phase, where ices evaporate 
from grain surfaces, and high-temperature chemistry is important.
They showed that the dynamical timescale in  Shu's inside-out collapse model
was too short compared to the chemical timescale to develop the observed 
molecular abundances in massive hot cores. 
Doty et al. (2004) also constructed a self-consistent radiative transfer model 
including dust radiative transfer, gas energetics, and a 
time-dependent chemical network to simulate molecular line transitions and,  
finally, to compare with observations in IRAS 16293-2422.  
They adopted a static protostellar envelope and assumed initial 
abundances based on observations of massive star-forming cores.
They did not include the freeze-out process in the chemistry 
and considered that only CO, H$_2$CO, and CH$_3$OH were desorbed
in the outer envelope that has temperatures below 100 K.  
     
In addition to these theoretical studies, Lee et al. (2003) and J{\o}rgensen et
al. (2004b) have tested some simple empirical abundance profiles, such as
a step function  and a drop function (see Figure 9 of J{\o}rgensen et al. 2004b) 
to fit observed line profiles of some molecules in the PPC stage and in the 
Class 0 stage, respectively. 

In this paper, we calculate the evolution of chemical abundances and line 
profiles of several important molecules, which have been used to study 
star-forming cores at submillimeter wavelengths, in a more self-consistent
fashion than seen in earlier work.
We include a specific dynamical model, the dust continuum 
radiative transfer, the gas energetics, a chemical model, and the line 
radiative transfer.
Our models include evolution in the PPC stage and the later evolutionary phase 
after collapse begins. 

In \S 2, we describe all models used for this study, and \S 3 and \S 4 show 
results of our modeling. We discuss in \S 5 how various molecular line 
tracers depend on evolutionary stage, cloud environment, etc.
Finally, we summarize our conclusions in \S 6.

\section{MODELS}

In this section, we describe our modeling process from dynamics to line profile
simulation through continuum radiative transfer, gas energetics, chemical 
evolution, and line radiative transfer.
The flow chart in Figure 1 shows the sequence of the modeling.
First, density and velocity fields [$n(r)$ and $u(r)$] as a function of 
time are provided by star formation theory. 
Most dynamical theories do not provide the temperature distribution [$T(r)$], 
but theories do usually provide a way to calculate the luminosity
of the central object versus time [$L(t)$]. 
Therefore, we can calculate the dust 
temperature distribution [$T_D(r)$] in the envelope at any point in the 
evolution by calculating the energy transfer through the dust. 
In this dust continuum radiative transfer calculation, we have to assume 
the opacity of dust grains ($\kappa_\nu$) and the gas to dust mass ratio 
as well as calculating the luminosity of the central object.
Once we calculate the dust temperature distribution, the gas temperature 
distribution [$T_K(r)$] can be computed from the gas energetics.
Deep in the cloud, where densities are high, collisions between gas and dust
maintain $T_K=T_D$, but, at lower densities, $T_K$ must be calculated by 
balancing heating and cooling rates.
In order to simulate molecular line profiles, we need to know the distributions
of molecular abundances, so we adapt the chemical model of Bergin et al. (1995)
to compute the chemical abundances as a function of radius [$X(r)$] at any
evolutionary time.
We calculate the abundance changes following every parcel of a model core 
through the whole evolutionary time in a Lagrangian coordinate system because
the abundances established in the outer layers are carried inward with the
matter.
After the calculation, all parcels are transferred to Eulerian coordinates 
so as to give the distributions of molecular abundance at any given time.   
The next step is to calculate level populations of molecules with the Monte 
Carlo (MC) method to solve the coupled excitation and the radiative transfer 
self-consistently.
The final step is to  simulate line profiles at each time step by calculating 
the intensity emitted by the model core at the distance of $d$ over the 
spectral line velocities at any impact parameter, convolving to match the 
spectral ($\delta v$) and spatial resolution ($\theta_{FWHM}$), and 
the main beam efficiency ($\eta_{mb}$) of the observations.  
Each step of the whole process of this modeling is described in more detail 
in following subsections.  

\subsection{Dynamical Model}

We use the inside-out collapse model (Shu 1977) as a test dynamical collapse 
model.  
However, we need to consider the initial conditions in the chemistry carefully 
because previous studies have shown that many molecules are depleted in dense 
cores (PPCs), whose central density is high enough to make a big contrast to 
their boundary density, but which do not contain a source of central luminosity.

The submillimeter continuum emission from those PPCs indicates that the density
profile does not follow a power law all the way to the center. Instead, it shows
roughly a flat density distribution within some radius. 
Such density profiles are
well described by Bonnor-Ebert spheres. Evans et al. (2001) showed that 
Bonnor-Ebert spheres fit the observed data well through detailed analysis of
the submillimeter emission from dust toward three PPCs (see also Alves, Lada, \& Lada 2001). 
The cores, whose only heating source is the interstellar radiation field (ISRF),
are very cold in the interiors ($T_D \sim$7 K).
As a result, the densities needed to match the observed emission are higher than
what had been thought previously. This is important for studies of molecular evolution
because higher densities enhance the freeze-out rate.

For higher central densities, the region of the flat density distribution in 
Bonnor-Ebert spheres is smaller, and the outer regions are similar to a power 
law ($n(r) \propto r^{-p}$) with $p \sim 2$.  
As the central condensation is growing, the spheres approach the singular 
state, which is used for the initial condition of inside-out collapse.
Therefore, we combine a sequence of Bonnor-Ebert spheres and inside-out 
collapse in order to set up a dynamical model to evolve from 
the pre-protostellar stage to later stages.
The outer radius for both Bonnor-Ebert spheres and inside-out 
collapse used in this work is 0.15 pc.
We assume a constant density structure of $10^3$ \cmv\ as the very first 
condition, and the model core is embedded in surroundings with some amount of 
visual extinction.
Figure 2 shows the density profiles of 7 Bonnor-Ebert spheres and the density
profiles of inside-out collapse in several time steps.      
The (7) Bonnor-Ebert evolutionary steps assume an evolution in central density 
from $10^4$ to $10^7$ \cmv. 
The Bonnor-Ebert spheres do not evolve dynamically, so there is no good
constraint on timescales in the PPC stage. 
Therefore, we do not know the exact timescale spent at each central density, 
but can expect that the timescale decreases as the central density increases.  
Based on the free-fall timescale ($\sim 1/\sqrt n$) and the 
statistical timescale of the whole PPC stage (see e.g. Lee \& Myers 1999),
we simply assume the timescales of $2.5\times10^5$,
$1.25\times10^5$, $6\times10^4$, $3\times10^4$, $2\times10^4$, $1\times10^4$,
and $5\times10^3$ for Bonnor-Ebert spheres of $n_c=10^4$, $3\times10^4$, $10^5$,
$3\times10^5$, $10^6$, $3\times10^6$, and $10^7$ cm$^{-3}$ respectively.
The timescale for the very initial constant density ($\rm n=10^3$ cm$^{-3}$)
is $5\times10^5$ years. 
Therefore, the total timescale for the PPC stage is a million years. 
We, however, expect to give some constraints to the timescales by comparing the
results of this work with actual observations. 
For instance, we know that the most important 
effect of the PPC stage is the freeze-out of
molecules, so the longer this stage lasts, the more the lower density 
in the outer layers will be affected.

We obtain density profiles from the Bonnor-Ebert sphere model and 
inside-out collapse model assuming that the model core is isothermal with 
the kinetic temperature of 10 K.
Evans et al. (2001) showed that differences between the isothermal Bonnor-Ebert
spheres and non-isothermal Bonnor-Ebert spheres are small.
The total core mass is 3.6 $\msun$.
The properties of the model core are summarized in Table 1.

In Bonnor-Ebert spheres, all gas parcels are in steady state, but the density 
at a given position increases as time goes by without any noticeable flow of 
gas.
In collapse, all gas parcels inside the infall radius move inward without
any interaction with adjacent gas parcels, but gas parcels outside the infall
radius stay at the original radii without any change of density distribution 
with time.
In order to study the chemical evolution in a model core, we should follow 
every gas parcel that experiences different physical environments with time 
(best done with a Lagrangian coordinate system) and calculate chemical 
reactions adopting the results of chemical evolution from one time step earlier 
as the initial condition at a given time step.
Finally, we transfer all gas parcels again to Eulerian coordinates for each 
given time step to determine the abundance profiles of molecules with time.  
We consider 190 time steps in the inside-out collapse to make total 198 
time steps for the whole evolution.
The time step varies with time. The velocity change slows down
with time, so a smaller time step has been used for earlier times;
500 years for the first 10,000 years, 1000 years from 10,000 to 100,000 years,
and 5000 years from 100,000 to 500,000 years.

We use the continuity equation in Lagrangian coordinates to realize this idea 
in practice.
The continuity equation in spherical coordinates is 
\begin{equation}
\frac{dn}{dt}=-n \nabla \cdot \overrightarrow{v}=-n \frac{1}{r^2}
\frac{\partial}{\partial r}[r^2 u]
=-n\frac{1}{r^2}[2ru+r^2\frac{\partial u}{\partial r}]
=-n[\frac{2u}{r}+\frac{\partial u}{\partial r}],
\end{equation}
where $\overrightarrow{v}$ is a velocity vector, and $u$ is the radial 
component of velocity. The change of density
of a parcel in a small time interval, $dt$ is given by 
\begin{equation}
dn=-n[\frac{2u}{r}+\frac{\partial u}{\partial r}]dt.
\end{equation}
Therefore, the density of a parcel in the next time step is given by
\begin{equation}
n(t_2)=n(t_1)+dn=n(t_1)-n(t_1)[\frac{2\overline{u}}{r}+
\overline{\frac{\partial u}{\partial r}}]dt, ~~~dt=t_2-t_1
\end{equation}
where $t_1$ and $t_2$ represent a given time step and the next time step,
respectively, and $\overline{u}$ and 
$\overline{\frac{\partial u}{\partial r}}$ is the mean velocity and mean 
velocity gradient between $t_1$ and $t_2$, respectively. 
We use mean velocity and mean velocity gradient because the velocity 
in collapse changes continuously with time, and the time interval for 
this work is not infinitesimal.
The other reason that we use the mean values is in the propagation of the 
infall radius outward at the speed of sound. 
The region between two infall radii at $t_1$ and $t_2$ has 
zero speed at $t_1$, so we cannot use $u(t_1)$ and 
$\frac{\partial u}{\partial r}(t_1)$ to calculate the change of density of 
a parcel.
We assume $dn=0$ if $dn$ has a negative value.
$dn$ can have a negative value between infall radii at $t_1$ and $t_2$ because
$u(t_1)=0$ at those radii so that $\overline{u}$ is a half of $u(t_2)$, and 
in turn, its absolute value can be smaller than that of 
$\overline{\frac{\partial u}{\partial r}}$, which is always positive.

Once we calculate $n(t_2)$, we can find the radius, temperature, and velocity
of the given parcel from the solutions of inside-out collapse at $t_2$. 
We follow all 512 gas parcels until they fall onto the central protostar.
We assume that a gas parcel falls onto the central protostar if the radius of
the parcel is less than the inner radius of the model, 70 AU.
The falling time depends on the initial position; that is, the smaller the 
radii gas parcels are at initially, the faster they disappear from the envelope.  
Figure 3 shows the evolution of density, temperature, radius, and $\rm A_V$ of 
several parcels with time. 
$\rm A_V$ is calculated from outside to a given parcel.
We consider the initiation of collapse as $t=0$, so the evolution is divided 
into $t<0$ and $t>0$ for the Bonnor-Ebert spheres and for the inside-out 
collapse, respectively.

\subsection{Dust Radiative Transfer}
In order to calculate dust temperature profiles ($T_D(r)$), we use a 1-d dust 
continuum radiative transfer code, DUSTY, created by Ivezi\'c et al. (1999).
PPCs are heated only by the ISRF (Evans et al. 2001). For $t>0$, we add a 
central luminous object.
Even for $t>0$, the ISRF significantly affects
both the total observed luminosity and the shape of the observed submillimeter
intensity profile (Shirley et al. 2002, Young et al. 2003).
Evans et al. (2001) found that observed data were better fitted with the ISRF 
reduced for wavelengths from the ultraviolet to far-infrared. 
We assume that the model
core is exposed to the Black-Draine ISRF (Evans et al. 2001) attenuated
by some amount of external visual extinction (A$_{\rm V,e}$) in the dust 
radiative transfer code because we test several external visual extinctions 
between 0.5 and 3 mag in chemical evolution. 
Our standard model for the chemical evolution adopts A$_{\rm V,e}=0.5$ mag. 
For the opacity, we use OH5 dust, which is found in column (5) of
Ossenkopf \& Henning (1994). This opacity, for grains that have coagulated
and accreted thin ice mantles, was found to match the multi-wavelength 
observations of low mass cores in all stages modeled here 
(Evans et al. 2001, Shirley et al. 2002, Young et al. 2003).
Figure 4 (left panel) shows $T_D(r)$ before collapse begins ($t<0$).
The dust temperature goes down to about 6 K at the core center because the 
model core is heated only by the ISRF at $t<0$.

For the evolution of the central source of  luminosity (for $t>0$), 
we adopt the model of Young \& Evans (2004).
After collapse begins, the central luminosity
source includes accretion from the envelope onto a central protostar and disk, 
the disk luminosity, and
the internal stellar luminosity due to contraction and deuterium burning.
Some of these sources turn on at different times.
At early times, the accretion luminosity dominates all other components.
Since there is no disk at very early times, accreting material falls
directly onto the central object.
The first hydrostatic core (FHSC) (Larson 1969, Boss 1993,
Masunaga et al. 1998), which has a large radius (5 AU), is included until
20,000 years. The transition between the large core to the smaller core occurs
between 20,000 and 52,000 years.
As the disk begins to form, most accreting material is deposited onto the
disk, so the accretion luminosity becomes the dominant luminosity source until
the photospheric luminosity becomes important at $t=100,000$.

Figure 4 (right panel) shows the dust temperature profiles calculated from 
the dust radiative transfer code for the model core after dynamical 
collapse begins.
In general, temperature increases with time at all radii as the central 
luminosity increases, and the envelope mass decreases, but there are two 
jumps between 2$\times 10^4$ and 5$\times 10^4$ and between $10^5$ and 
$1.6\times 10^5$ years.
The first jump is caused by the transition from the FHSC to the smaller core, 
and the second jump is due to the turn-on of the photospheric luminosity from 
gravitational contraction and deuterium burning in the central star at
$t=1\times 10^5$ years. 
Around 100000 years, the transition between Class 0 and Class I occurs, when
the bolometric temperature is about 70 K.

\subsection{Gas Energetics}

The gas kinetic temperature ($T_K$) is determined by the balance 
between heating and cooling. The gas heating at large densities is dominated by 
gas-grain collisions, but grain photoelectric (PE) heating 
is significant at the outer part of a core. Cosmic-ray (CR) heating
is also important in a transition region where PE heating has decreased and 
gas-grain collisions are not yet dominant.
Therefore, we use a new code updated by S. Doty to add PE and CR heating to the 
code of Doty \& Neufeld (1997), which considered only gas-grain collisional 
heating and adopted the cooling functions of Neufeld, Lepp, \& Melnick (1995) 
to determine the gas cooling rates due to emission from CO, O$_2$, O I, H$_2$, 
H$_2$O as well as from other diatomic and polyatomic species.
They computed the steady-state abundances of the atoms and molecules in pure
gas chemistry including the full UMIST chemical network (Millar et al. 1991). 

However, molecules such as CO (the dominant coolant in cold dark clouds)
are frozen onto dust grain surfaces at very 
dense and cold inner region of PPCs. 
In addition, once a central star forms dust grains can 
be heated above 100 K, where the evaporation of water from icy grain 
mantles occurs. Water is a very efficient coolant in hot regions. 
Therefore, one should, in principle, include the effects of freeze-out and
desorption to compute the 
correct abundances of coolants in the dense inner region, and iterate the
calculation of heating and cooling to find the converged gas temperature 
for different densities and dust temperatures.
However, Goldsmith (2001) and Doty \& Neufeld (1997) found the effects of 
depletion of CO and evaporation of water were relatively 
small at high densities because of efficient gas-dust coupling.
We tested the effect of CO depletion by reducing X(CO) by 2 orders of magnitude
in the gas energetics code.
The calculated gas temperatures are the same as the dust temperatures if the 
density is higher than about $10^5$ cm$^{-3}$, where the freeze-out of CO
becomes significant. 
In addition, the inner radius in our models is not small enough to reach
layers where the dust 
temperature is $>$100 K.
Therefore, we do not include the effects of freeze-out and desorption in the 
gas energetics.
Other possible inputs into the gas energetics are H$_2$
formation, FUV pumping of H$_2$, and H$_2$ dissociation heating. 
In our calculations we have assumed a weak attenuated UV field leading
H$_2$ to be fully shielded well before the layers examined by our
models.  Therefore the heating associated with H$_2$ formation,
dissociation, and FUV pumping will be small compared to photo-electric
heating and can be ignored (see Sternberg
\& Dalgarno 1989; Burton, Hollenbach, \& Tielens 1990). 

The gas energetics code can examine solutions with an attenuated UV field.
The CO lines, which are easily thermalized, around PPCs show much 
weaker emission lines than those expected if the cores are heated by
the unshielded local ISRF (K. Young et al. 2004).
In order to calculate $G_0$, which is the strength of the UV field relative
to the local ISRF at the outer cloud boundary, we use the equation, 
$G_0=exp(-1.8\times {\rm A_{V,e}})$. 
Figure 5 shows gas temperature profiles computed with this gas energetics code.
The gas temperature gets high in the outer regions because of photoelectric
heating. However, gas-grain collisions are dominant in dense regions, so the
gas temperature is well coupled with the dust temperature at small radii.
If the ISRF suffers less attenuation, then it heats gas better through the core 
to keep the gas temperature higher, especially, in less dense regions. 

Details of the gas energetics code used in this study are given in the appendix
of K. Young et al. (2004) and Doty \& Neufeld (1997).

\subsection{Chemical Model}

We use the time-dependent chemical model of Bergin, Langer, \& Goldsmith (1995),
updated in Bergin \& Langer (1997),
which includes depletion and desorption of atoms and molecules onto and off 
grain surfaces, as well as gas-phase interactions. However, the model does not 
explicitly include chemical reactions on grain mantles.  
To reduce the computation time we have used a reduced network including $\sim$80
species and $\sim$800 reactions.  
The reduced network was created by an examination of the major 
formation and destruction paths for key species.  
This was tested in comparison to the larger network.

As described in \S 2.1, chemistry is calculated following each 
parcel, which experiences different density and temperature environments as it 
falls toward the center. 
We consider 198 dynamical time steps from the PPC stage to later stages 
after collapse begins as mentioned in \S 2.1. 
In each time step, physical conditions are constant, so the chemistry
is calculated pseudo-time-dependently between two time steps by adopting the 
chemical results of the previous time step as the initial conditions in the 
later time step without changing physical conditions.

This model for gas-phase chemistry includes the photodissociation and 
photoionization of molecules as well as the cosmic ray ionization. 
For the cosmic ray ionization rate we have assumed $\zeta_{H_2}$ = 1.2 $\times$
10$^{-17}$ s$^{-1}$ and the abundance of He$^{+}$, a key destroyer of many 
species, is derived from a balance of cosmic ray ionization of He 
($\zeta_{He^+}$ = 6.5 $\times$10$^{-18}$ s$^{-1}$)
and its major reactants (especially CO).
However, to reduce further the computational complexity 
we have not included CO photodissociation in the model and only allow
its abundance to be altered via gas-phase and gas-grain interactions. 
To minimize the effects of this on the chemical calculation we have assumed
there exists additional extinction (A$_{\rm V,e}$) beyond the edges of 
our model.
CO is self shielding for Av between 0.5 and 1 mag, so most of our results are 
unaffected by neglecting CO photodissociation. 
In our models we have examined external extinctions between 0.5 -- 3 mag.  
At its lowest value  the abundances of simple radicals (e.g. CN, C$_2$H) that
form in the outer photo-dissociation layers could be affected.

For the gas-grain reactions, this model assume that grains are charged with 
one electron per grain, and positive ions undergo dissociative recombination
with the same branching ratio as in the gas phase when they collide with
grains.
This model also considers three important desorption mechanisms: thermal 
evaporation, cosmic-ray-induced heating, and direct photo-desorption 
(see Bergin, Langer, \& Goldsmith 1995 for details). 
At low temperatures as in PPCs, 
cosmic-ray-induced desorption is dominant. Once collapse begins, however,
$T_D$ increases, so thermal evaporation becomes important.
Freeze-out of molecules is highly dependent on the dominant component of 
the grain mantle, that is, the surface binding energy of the species. 
We use the surface binding energy of species onto a bare SiO$_2$ dust
grain as a standard. Table 2 lists the binding energies of selected
species. 
We also test binding energies onto H$_2$O and CO dominant grain mantles.
We assume a sticking coefficient of 1.0 for all species.
In order to reduce the water abundance in the gas to match SWAS observations 
(Bergin et al. 2000), we adopt a binding energy of 
5700 K (Fraser et al. 2001) for water onto grain surfaces. This binding energy is the same
even on different dust grains, unlike the binding energies of other molecules.
In addition, we consider that atomic oxygen is frozen onto grain surfaces
in the form H$_2$O on grains again to match SWAS observations.
For photodesorption, this model uses the mean 
interstellar UV field (Habing 1968), which is attenuated as a parcel moves 
inward based on the calculated $\rm A_V$. 
We have used the Black-Draine field for the dust heating calculations and the 
Habing field for all calculations of photodissociation and photodesorption.
This small, factor of 1.7, inconsistency will not affect our results 
given uncertainties in photodissociation rates and photodesorption
yields.
We assume that the core evolution occurs in a shielded
region, with different external extinctions to represent various 
environments of star-forming cores.
The initial elemental abundances are listed in Table 3.
Initially (t$=-10^{6}$ years), all hydrogen is molecular (H$_2$), and all 
carbon is ionic (C$^+$).  

For more detailed explanation of this chemical model, refer to Bergin et al. 
(1995). 

\subsection{Molecular Radiative Transfer Model}

Once all physical conditions [$n(r),~T_K(r),~u(r),~X(r)$] are computed from 
other models we perform a radiative transfer calculation for the spectral 
lines.
For the molecular radiative transfer calculation, we use the MC code that  has
been developed by Choi et al. (1995). The MC code generates model photons at a
random position, in a random direction, and at a random frequency with proper
random number distributions. The MC code calculates the excitation by the 
model photons and uses statistical equilibrium to adjust each level population 
until the criteria of convergence are satisfied.  
We can model arbitrary distributions of systematic velocity, density, 
kinetic temperature, microturbulence, and abundance self-consistently with the
MC code. We use collision rates calculated  by Green (1992) for CS, by Flower 
\& Launay (1985) for CO, by Green (1991) for H$_2$CO, by Flower (1999) for 
HCO$^+$, and by Turner (1995) for $\rm N_2H^+$. 
Once the MC code calculates each level population, we simulate specific 
molecular line profiles emitted by the model core at the assumed distance, $d$, 
over the spectral line velocities at any impact parameter by calculating the
radiative transfer along rays and by convolving to match the spectral and 
spatial resolution of the observations with a virtual telescope simulation 
program. 
For this study, we adopt the distance of Taurus, 140 pc, and spatial resolutions
of several telescopes that we have used for studying star-forming cores. 
The simulation program can model molecular lines with multiple transitions and 
multiple impact parameters simultaneously.

We model lines of CO, C$^{18}$O, CS, HCO$^+$, and $\rm N_2H^+$ 
at submillimeter wavelengths.
Those lines have been observed with the 10-m telescope at the Caltech 
Submillimeter Observatory, the 15-m James Clerk Maxwell Telescope, and the 45-m 
telescope at the Nobeyama Radio Observatory toward PPCs, Class 0 and I objects 
(Lee et al.  2003; Lee et al. 2004b).
We also model the ortho-H$_2$CO 6 cm line, which has been observed at Arecibo 
Observatory (K. Young et al. 2004).
Therefore, the simulation of molecular line profiles and comparison with
the actual data in different evolutionary stages will provide a possible 
opportunity to test chemistry as a tracer of star formation.  
The comparisons will appear in separate papers, as mentioned above.

\subsection{Characteristic Timescales}
We compare characteristic timescales for this model in Table 4 for a
density of $10^5$ cm$^{-3}$.
We assume steady sate for the calculation of dust temperature, gas 
temperature, and the level population of molecules.
The main heating source of the dust energetics is the radiative heating. 
The energy diffusion timescale inside grains
is much smaller than the photon absorption timescale, i.e., the interval
between consecutive arrival of photons. Therefore, the timescale of the
dust energetics is mainly dependent on the photon absorption timescale, which
is very short compared to dynamical and chemical timescales. 
In this work, we neglect small, transiently heated grains because they are not
important to the processes that we consider in well-shielded regions, such as 
dense molecular cores (Bernard et al. 1993).
The timescale for the gas temperature to equilibrate with the dust
temperature through dust-gas collisions is also short enough to be ignored when
we consider chemical and dynamical evolution in the high density regions.
The timescale for molecular excitation is also much shorter than
chemical and dynamical evolution (see Table 4), supporting the
steady state approximation for the excitation of molecular levels.

We have to consider dynamics and chemistry simultaneously once collapse begins 
($t>0$) because their timescales are similar within the infall radius.
The exact process during the PPC stage is still a matter of debate, and cores 
may coalesce either through dissipation of turbulence (Myers \& Lazarian 1998) 
or ambi-polar diffusion (Ciolek \& Mouschovias 1993).
In this work, for $t<0$, we assume a sequence of Bonnor-Ebert spheres with a 
total timescale fixed at $10^6$ years with shorter timescales for denser stages.
Although some depletion is expected during the early stages,
the largest effect on the chemistry will occur at the highest densities. 
Therefore, the relevant timescale for the PPC stage is the time spent
at the greatest density prior to collapse.
The evaporation timescale depends exponentially on $T_D$. This timescale
is either extremely long ($T_D < T_{evap}$) and evaporation can be ignored, or
it is extremely fast ($T_D > T_{evap}$) and can be considered to be 
instantaneous. The evaporation process yields molecules that drive gas phase 
chemistry with its longer timescale.
From Table 4 and this discussion, it is clear that we need to treat
the dynamical and chemical evolution together in a time dependent way,
while we can treat the dust and gas energetics and molecular excitation
in the steady state approximation.

\section{RESULTS OF CHEMICAL MODELS}
\subsection{Standard Model}
The results of our standard chemical model, where the ISRF is attenuated by
A$_{\rm V,e}=0.5$ mag, and silicate binding energies are used for
gas-grain reactions, are shown in Figures 6$-$11.
In the following we will separate the terms freeze-out and depletion which are
commonly used to reference similar effects.  Here we use the term 
freeze-out to refer to the removal 
of molecules from the gas and onto grain surfaces.   Depletion is used
in reference to the decrease in abundance of molecular ions, which are believed
to not freeze-out directly onto grains.  Ions will decrease in
abundance, or deplete, when a parent species freezes onto grain surfaces
(e.g. CO for HCO$^{+}$). 

\subsubsection{Time Dependance}
Figure 6 shows the evolution of molecular abundances in gas at given radii. 
The timescale on the x-axis starts from $-5\times10^5$, when CO reaches around 
its normal abundance of $10^{-4}$. 
We will concentrate at first on the evolution in the central r$=$0.001 pc.
The abundances of molecules increase at very early times, and decrease
as freeze-out/depletion acts at $t\leq 0$; then at $t> 0$, heating by the 
central source increases the dust temperature to the evaporation temperatures 
of molecules.\footnote{The evaporation temperature ($T_{evap}$) is 
$T_D$ when the evaporation timescale 
($\tau_{evap}$) is equal to the freeze-out timescale ($\tau_{freeze-out}$).}

After 50,000 years the dust temperature  
reaches the CO evaporation temperature ($\sim 25$ K), so the CO abundance 
quickly increases.
The sharp increase of CO abundance also 
enhances the formation rates of CS, $\rm H_2CO$, and HCN.
This is more readily shown in Figure 7 where the time dependence of CS and 
$\rm H_2CO$ is highlighted.  Here the CS and $\rm H_2CO$ gas abundances
show the increase at $\sim$50,000 years, but the ice abundance is still
increasing. In the gas the formation of CS, $\rm H_2CO$, and HCN is then 
indirectly powered by reactions of evaporating CO with H$_3^+$ and He$^+$. 
The He$^+$ reaction in particular creates the fuel 
to power other chemistries. The CS chemistry is aided by the similar evaporation
temperature of CO and sulfur atoms.

As time increases the increasing dust temperature approaches 
the evaporation temperature of the more tightly bound CS, $\rm H_2CO$, and HCN.
Upon evaporation the gas-phase abundance of HCN and $\rm H_2CO$ is higher than
can be sustained via chemical equilibrium and the abundances drop.
The primary process that limits the abundances of newly evaporated species
is reactions with H$_3^+$, and to a lesser extent He$^+$.
{\em Thus the presence of dust temperatures above the evaporation point does
not necessarily dictate high abundances.}
However, CS shows a different evolution as after evaporation its abundance does
not drop.  This is due to the freeze-out of atomic oxygen and water at small 
radii. 
If they are not frozen onto grain surfaces, the sulfur goes to SO over CS at 
late times in the chemical evolution. However, they exhibit significant freeze-out
in the central region, so CS stays as the main sulfur-bearing species.    
As a result, the CS abundance at the later time becomes greater than the normal
abundance.

The abundances of  $\rm HCO^+$, $\rm NH_3$ and $\rm N_2H^+$ deserve special attention.
The chemistry of $\rm HCO^+$ is rather simple as it tends to follow that of CO.
$\rm NH_3$ and $\rm N_2H^+$ are also influenced by CO but instead show abundance
increases as CO freezes onto grains.
Thus, at r$=0.001$ pc, all molecules except for $\rm NH_3$ and $\rm N_2H^+$ 
deplete or freeze-out until 50,000 years.  The sharp peak of $\rm NH_3$ at this time is 
due to its evaporation.   Similar to the other species the gas phase equilibrium
cannot sustain the high abundances from the evaporated  $\rm NH_3$ and its
abundance drops.
In the case of $\rm N_2H^+$, CO is a major destroyer and the abundance of 
$\rm N_2H^+$
decreases sharply in reaction to the CO abundance increase around 50,000 years.

At large radii, the dust temperature reaches the evaporation temperature at
timescales later 
than at smaller radii, so the duration of freeze-out is longer. However, the
amount of freeze-out is smaller because of the lower density, thus longer
freeze-out timescale.
The evaporation peaks of $\rm H_2CO$, HCN, and NH$_3$ disappear at large
radii because the dust temperature never reaches their evaporation temperatures.
After CO evaporates, $\rm N_2H^+$ is more abundant
at large radii due to the higher abundance of H$_3^+$ at lower densities.

\subsubsection{Radial Dependance}

In Figure 8 and 9 we present the radial abundance profiles of CO and other 
molecules in the gas phase at selected times.\footnote{Movies of the evolution 
of abundance profiles as well as density, temperature, and molecular line 
profiles for all time steps are found in 
{\bf http://peggysue.as.utexas.edu/sf/CHEM/}.}
The left boxes of the Figures represent abundances in gas before collapse 
begins. The CO abundance (Figure 8a) in the gas phase increases at 
$t\le -2.5\times 10^5$ years. At later times we see the significant effect of
freeze-out which has a strong dependence on the radius because the freeze-out 
timescale increases going outward.
When, however, $T_D$ reaches the CO evaporation temperature, 
we see a large abundance rise in the center.
Our chemical model includes only one molecular species, H$_2$ at the
beginning of evolution.
CO reaches the abundances of $10^{-5}$ and $10^{-4}$ at $-9.8\times10^5$ and
$-4.7\times10^5$ years, respectively, at the center.
Once $T_D$ reaches $T_{evap}$ of CO (around $t=50,000$ years at $\rm r=0.001$
pc), almost all CO in the ice phase evaporates, so a sharp evaporation front
is produced.
This evaporation front propagates outward as the dust temperature grows with
time.
The gas-phase CO abundance increases gradually with radius to make a dip, 
where the CO freeze-out is the most significant, just behind the sharp 
evaporation front. 
This is because the freeze-out timescale increases with radius as shown in
Figure 10.
For r $\geq$ 0.08 pc (A$_{\rm V}\leq$ 1), photodesorption will remove CO 
from grain mantles.
As the result of the CO chemistry depending on the dust temperature and density,
one can simplify the abundance profile to three zones; evaporation,
freeze-out, and normal abundance.
J{\o}rgensen et al. (2004b) used a simple ``drop" function, which reasonably 
approximates this profile, to simulate C$^{18}$O lines to fit actual
observations.

The evolution of the abundance profile of H$_3^+$, which 
initiates the chemical pathways of other molecules, is shown in Figure 8b.
It decreases toward the center in all time steps because of the increase of 
density toward the center. The H$_3^+$ abundance decreases sharply at radii
smaller than the CO evaporation front because H$_3^+$ reacts with the 
evaporated CO. 
In Figure 8c, the thick dot - short dash and thick solid lines compare the 
abundance profiles of CO and H$_3^+$ at t$=10^5$ years. The long dashed line
indicates the H$_3^+$ abundance profile if the H$_3^+$ density is constant
through the core. This is a usual condition assumed in molecular clouds
(Geballe et al. 1999). 
As shown, however, only at radii smaller than the CO evaporation 
front do the model results follow the condition of a constant
H$_3^+$ density.
The H$_3^+$ abundance profile is significantly affected by the distribution 
of CO freeze-out at radii greater than the CO evaporation front.
In addition, the H$_3^+$ abundance increases with time at almost all radii.
This results from the combination of dynamical and chemical evolution. 
Once the gas and dust are within the infall radius the higher abundances
established in lower densities are carried inward.
For instance, the peak at r$=0.08$ pc stays at the same
position at t$=10^5$ and $2\times 10^5$ years because the infall radius has
not reached the position. However, at t$=5\times 10^5$ years, the infall
radius is greater than the radius of the position, so the peak moves inward
to r$\sim 0.05$ pc. This effect is shown better in Figure 9 if we compare
abundance profiles at t$=2\times 10^5$ and $5\times 10^5$ years.

CS (Figure 9) shows the same trend as CO, i.e., CS is removed 
from the gas phase more with time until $T_D$ reaches the 
$T_{evap}$ of CO.
After $T_D$ reaches $T_{evap}$ of CS, the abundance profiles are more 
complicated than those of CO. 
As a general trend, there are two jumps in the CS profiles (shown better in
Figure 10) in the inner region, where dust-gas chemistry is important.
One is at the radius of $T_{evap}$ of CS, and the other is coincident with
the evaporation front of CO.     
The first jump can be understood in the same way as CO, but the second 
is related to the CO evaporation as mentioned in the previous section. 
At radii smaller than the first peak, the abundance stays constant, 
greater than the unfrozen, normal abundance at outer radii. 
At radii greater than the second peak, the freeze-out dip follows, and the CS
abundance increases to reach its normal abundance. 
However, at radii greater than around 0.02 pc, CS is dissociated by UV photons.

H$_2$CO (Figure 9) shows similarities to both CS and CO, and is frozen on grain 
surfaces before collapse begins.
Once $T_D$ reaches its evaporation temperature,
the abundance profile is very complex. 
Figure 10 shows the H$_2$CO abundance profiles in two time steps,
and $\tau_{evap}$ and $\tau_{freeze-out}$ in the previous time step.
There are two peaks in the abundance profile. 
The biggest peak around 0.001 pc is due to the evaporation of the 
H$_2$CO ice built up in the earlier stage.
At radii smaller than 0.001 pc, the H$_2$CO abundance decreases toward
the center, which is due to the equilibrium gas phase chemistry being
incapable of sustaining the high abundances from the evaporated ice. 
The peak around 0.003 pc, overlaps the CO evaporation front,
and is related to the sharp increase of the CO abundance.
At radii greater than 0.003 pc, the abundance increases with radius
because the freeze-out timescale increases with radius, so H$_2$CO is less 
frozen-out.

HCN shows a trend similar to that of H$_2$CO, which has two peaks at the radii 
of CO and H$_2$CO evaporation temperatures. 
However, the HCN profile shows even more structure.
At any given time HCN shows three abundance peaks.  One is due to 
CO evaporation and the other HCN evaporation.  The third is related to the 
release of CN from grain mantles (CN has an estimated binding energy between
those of HCN and CO). Gas-phase reactions between H$_3^+$ and CN result in the
production of HCN. 
At 50,000 years, where there is the transition between the first hydrostatic
cores and Class 0 objects, the abundance of HCN increases toward the center by 
about a factor of 10 so that the abundance 
at all inner radii is comparable to the normal abundance in spite of structure.  
This is due to the evaporation of CO reaching its peak where all CO is 
returned to the gas.  At this time reactions of CO with He$^+$ ions reach
the peak destructive power and a small amount of carbon is transferred to HCN.
Therefore, HCN could be a good probe to distinguish the Class 0 stage from
the first hydrostatic core stage.

NH$_3$ is a molecule formed at later evolutionary stages, 
so it becomes more abundant with time for $t<0$.
However, it also freezes out in the center before the dust 
temperature reaches its evaporation temperature, which is assumed to be 
slightly lower than that of CO. Figure 10 shows that the evaporation front of 
NH$_3$ is located just behind the CO evaporation front. 
This small difference of the assumed evaporation temperatures of two molecules causes 
a peak between two fronts and sharp drop of the NH$_3$ abundance at smaller 
radii than the CO evaporation front as the equilibrium cannot maintain 
the high abundance of recently evaporated NH3.

HCO$^+$ (Figure 9) is formed by the reaction between CO and H$_3^+$, so it 
tends to follow CO.
However, unlike CO, HCO$^+$ shows a decrease at radii smaller than the 
CO evaporation front.
This trend is related to the evolution of H$_3^+$ which decreases in 
abundance at the small radii (for a given time) because of the high densities. 
In addition, HCO$^+$ increases at all radii with time because
the infalling material carries the H$_3^+$ and HCO$^+$ abundances 
established in lower density outer regions to smaller radii.

$\rm N_2H^+$ (Figure 9) is formed from the interaction between H$_3^+$ and 
N$_2$, which is a molecule formed at later evolutionary stages and has a lower 
binding energy, so it becomes abundant later and depleted less than other 
molecules such as CO, CS, H$_2$CO, and HCN. 
$\rm N_2H^+$ is mainly destroyed by CO in dense regions so that it has an 
abundance hole inside the CO evaporation front at $t>0$.
This relation between CO and $\rm N_2H^+$ has been discussed by 
J{\o}rgensen et al. (2004a) and Lee et al. (2004a). 
The abundance of $\rm N_2H^+$ increases at all radii, except for r$>0.01$ pc 
at t$=5\times 10^5$ years in Figure 9, with time, like HCO$^+$. 

In Figure 9, the abundance of all molecules drop at radii greater than around 
0.01 pc from 200,000 years (magenta) to 500,000 years (orange).
The drop is caused by gas with a different chemical state being carried inward.
The abundance peaks around 0.03 pc do not move until t$=2\times 
10^5$ years because the infall radius is smaller than 0.03 pc at this time.
However, the infall radius is $\sim$0.11 pc at t$=5\times 10^5$ years, so
the peaks move inward to r$\sim 0.01$ pc. At this moment, the CO evaporation
front is also around 0.01 pc producing sharp drops of the abundances of HCN,
NH$_3$, and N$_2$H$^+$.

Figure 11 compares abundance profiles of seven different molecules at a given
time ($t=100,000$ years) to illustrate the importance of the CO chemistry. 
There are two big bumps in the profiles of CS, H$_2$CO, and HCN (for the moment
ignoring the rise due to CN evaporation) 
The first one is caused by the evaporation of the 
molecules from grain surfaces.
The second one is located at the same radius as the CO evaporation front.
The chemistry of those molecules is affected by the sharp increase of
the CO abundance.
The abundance profiles of NH$_3$, $\rm N_2H^+$, and HCO$^+$ are mainly dependent
on that of CO because CO is the major destroyer of NH$_3$ (indirectly) and 
$\rm N_2H^+$ (directly), but the main supplier of HCO$^+$.

Rawlings et al. (1992) used the inside-out collapse model to examine the 
chemical evolution at every point in the envelope. They included freeze-out 
without considering desorption to conclude that molecular ions should be good 
tracers of collapsing envelopes because the ion destruction rate decreases as 
the neutral species are removed from the gas phase onto grain surfaces. 
However, Choi et al. (1995)  did not find any evidence for the freeze-out of CS
in B335.
Our models with desorption can explain the abundant CS during protostellar
collapse ($t>0$) as we described above and in \S 3.1.1. 
In addition, based on the narrowness of emission lines in B335 (Menten et al. 
1984), Rawlings et al. (1992) suggested that the NH$_3$ lines could not trace 
high velocity infalling material due to the freeze-out. 
However, we have found that the drop of the NH$_3$ abundance, at small radii
at $t>0$, is caused from the CO evaporation rather than the freeze-out of 
NH$_3$ itself.  
Therefore, desorption should not be ignored even in low-mass star-forming 
regions.

We note that we have not included any grain surface chemistry beyond 
the production of gas-phase H$_2$ and H$_2$O ice.  
Some molecules such as NH$_3$, H$_2$CO, CH$_4$, and CH$_3$OH could potentially 
form via grain surface reactions and, if so, their abundances in the inner 
"hot core" could be enhanced in the immediate evaporation states.

\subsection{Other Models}
Figure 12 and 13 show abundance profiles modeled with three different external 
visual extinctions, at the initiation of collapse ($t=0$) and $t=100,000$ years.
Note that the edge of the cloud in Figure 12 and 13 is not directly exposed 
to the ISRF due to A$_{\rm V,e}$.
The main differences are higher abundances at the outer radii for higher 
$\rm A_{V,e}$, caused by lower photodissociation 
of molecules by the interstellar UV photons or lower dissociative recombination
with electrons.
However, CO does not show any difference among different A$_{\rm V,e}$ 
because CO photodissociation is not included.
CS is less abundant even at inner radii as well as outer radii 
in the standard model with $\rm A_{V,e}=0.5$ mag than in the models of higher 
$\rm A_{V,e}$ at $t=0$ (Figure 12).
This is due to the low extinction in the early, pre-freeze-out, Bonnor-Ebert 
evolutionary phases, which favors CS formation over SO.
At such early times models with higher extinction produce more SO which
freezes onto grain surfaces.
In the standard model, CS is more abundant at the radii smaller than 0.0025 pc
at 100,000 years (Figure 13).
The abundance difference at inner radii is the biggest between 0.001 pc 
and 0.002 pc.
In the standard model, the CS abundance does not change much (by a factor of 2),
but it varies by the 2 orders of magnitude in the model with $\rm A_{V,e}=3$ 
mag.  This large effect is due to the difference in the overall evolution of sulfur pool.
In the standard model there is higher amount of S atoms frozen on grains
which is seen in a reduction of gas-phase CS abundance in Figure 12.  Sulphur
atoms are assumed to evaporate at similar temperatures to CO and thus the 
sudden return of sulfur and carbon to the gas fuels CS production.  
For the higher extinction cases there is a large amount of SO on grains which 
when evaporated does not lead to efficient CS creation. 
The CS abundance at the cloud edge for the model with A$_{\rm V,e} = 0.5$ mag 
does not show a rapid decline from photodissociation due to a higher abundance 
of ionized carbon at r=0.01 pc which results in efficient CS formation.
The ionized carbon exists due to slower chemical timescales in the 
outer layers.

Figure 14 and 15 compare results of models with different binding energies in 
dust grain reactions.
The binding energies to the CO mantle and the H$_2$O mantle are smaller and 
greater than that to the SiO$_2$ mantle by the ratio of 0.82 and 1.47, 
respectively.
At $t=0$, the model with higher binding energy exhibits greater 
freeze-out of CO, CS, and HCN onto dust grains and greater depletion of HCO$^+$.
Even NH$_3$ and $\rm N_2H^+$ show significant freeze-out/depletion in the model with higher binding energy  at this early evolutionary stage. 
The results of the model with the higher binding energy is much closer to the 
standard model results for CO, CS, HCN, and HCO$^+$, while the opposite is
seen for NH$_3$ and $\rm N_2H^+$.
Another difference is that HCN shows little freeze-out in the model 
with low binding energy.
Therefore, we can rule out some dust properties from the PPC observations by
comparing HCN with NH$_3$ and $\rm N_2H^+$.  
For example, if no freeze-out/depletion is seen in these three molecules, dust 
grains are covered by CO.
On the other hand, dust grains have water dominant mantles if all three 
molecules show freeze-out/depletion.  
If only HCN shows freeze-out, bare SiO$_2$ grains are the main dust component. 

After collapse begins (t $= 10^5$ yr case), 
the dust temperature increases, and the difference 
between various binding energies is more significant and complicated. 
Greater freeze-out of CO, CS, and HCO$^+$ depletion is observed 
in the model with the higher binding energy. 
In addition, the evaporation fronts of molecules are located at smaller
radii due to higher evaporation temperatures.
In contrast to the case at t$=$0, the results of the model with lower
binding energy are closer to those of the standard model in CO, CS, HCN, and
HCO$^+$.   
Sharp differences are also seen for 
NH$_3$ and $\rm N_2H^+$ where the model with high binding energy shows
significant abundance structure not seen in the other cases.
This structure is caused by the freeze-out of N$_2$ onto H$_2$O mantles.
Thus the abundance structure for NH$_3$ and $\rm N_2H^+$ is due to the  
evaporation of N$_2$ from the grains.
Models with lower binding energy have little N$_2$ freeze-out and thus
both NH$_3$ and $\rm N_2H^+$ are not directly affected by N$_2$ evaporation.

Bergin \& Langer (1997) used the same chemical code to study the chemistry in
the PPC stage. They showed that CO and HCO$^+$ were not depleted on CO grain
surfaces, unlike our result.
This discrepancy is caused by the different density evolution.
In their model, the density does not increase above $10^5$ cm$^{-3}$, but it
goes up to $10^7$ cm$^{-3}$ at the center in our model.
The higher density reduces the freeze-out timescales of CO, so CO
freezes-out even on CO dominant grain mantles, and HCO$^+$ depletes from the
gas phase. 

\section{RESULTS OF LINE SIMULATIONS}
We use a 30$\as$ beam, which is the typical beam size of the CSO 10 m telescope
around 230 GHz, for CO $2-1$, and HCO$^+$ $3-2$ lines.
The CS $2-1$ and HCO$^+$ $1-0$ lines have also been simulated with a 30$\as$ 
beam. 
For the $\rm N_2H^+$ $1-0$ line, the beam size (17$\as$) of the NRO 45 m 
telescope at 93 GHz has been used. 
The beam sizes for the C$^{18}$O lines have been adopted from J{\o}rgensen et al. 
(2002); 33$\as$, 21$\as$, and 14$\as$ for $1-0$, $2-1$, and $3-2$, respectively.
We assume that the ortho-H$_2$CO 6 cm line is observed with the Arecibo 
telescope ($\theta_b=60\as$). 
The HCN $1-0$ line is also simulated with various resolutions.
We use the same velocity resolution of 0.1 km s$^{-1}$ and
the same microturbulence of 0.15 km s$^{-1}$ for all lines.

Figure 16 shows the evolution of the CO $2-1$ and the CS $2-1$ lines. 
In our models the exclusion of CO photodissociation may modify the results with 
respect to line profiles.  However, the additional extinction included in the 
model limits these effects.
The CO $2-1$ line of the model with $\rm A_{V,e}=3$ mag is much weaker
than that of the model with $\rm A_{V,e}=0.5$ mag.
The CO $2-1$ transition has low critical density, so it can be thermalized
even at the outer radii, and the kinetic temperature in the former is much lower
than that in the latter at large radii as shown in Figure 5.
In addition, The CO $2-1$ line of the model with $\rm A_{V,e}=3$ mag shows
the blue asymmetry caused by infall.
However, The CO $2-1$ line of the model with $\rm A_{V,e}=0.5$ mag does not 
change much in all time steps except in the time step of 500,000 years.
The line at 500,000 years shows the inverted asymmetry rather than the blue
asymmetry. 
This is caused by the distribution of its excitation temperature.
The excitation temperature in the model with $\rm A_{V,e}=0.5$ mag
increases at radii greater than 0.016 pc (see Figure 17), causing the inverted
asymmetry.
This result suggests that we have to consider simultaneously parameters such as 
proper kinetic temperature and abundance profiles as well as density and 
velocity structures.

The CS $2-1$ line does not show a big difference in intensity between two 
models while the blue asymmetry profile is more prominent in the model with 
$\rm A_{V,e}=3$ mag.  
The similarity in intensity may be caused by the higher critical density of 
about $10^5$ $\rm cm^{-3}$ of the line, so lines are emitted dominantly from 
the inner, denser region, where the excitation temperatures of two models are 
the same.
However, the more prominent blue asymmetry in the model with $\rm A_{V,e}=3$ mag
can be caused by the higher CS abundance in the outer part in the model with 
the higher $\rm A_{V,e}$. 
The wing of the line becomes broader with time because the infall velocity 
increases with time.
Since the material of the model core falls onto the central object, the
density within the infall radius decreases with time, and the peak radiation
temperature of the CS $2-1$ line becomes smaller.

Figure 18 illustrates the importance of the CO chemistry, which causes a peak
in the CS abundance profile, even in line profiles.
If the second bump of the CS abundance profile at 100,000 years, which is caused
by the CO evaporation, is ignored, then the CS $2-1$ becomes weaker, and
the wing structure disappears.
Therefore, the wing structure in the CS $2-1$ line profile could be evidence 
that the dust temperature is high enough to evaporate CO or CS.
However, this wing structure could be confused by an outflow, which is
not considered in this work, so high spatial resolution maps are important. 

Figure 19 shows the evolution of HCO$^+$ $1-0$ and $3-2$ lines.
The lines become broader with time due to greater infall velocity at later
times.
The blue asymmetry profile, which is an indicator of the infall motion,
is robust in both models with two different external extinction environments.
In addition, the blue asymmetry of the $1-0$ and $3-2$ lines grow with time, 
as seen in Gregersen et al. (1997).
Both $1-0$ and $3-2$ lines of the standard model are weaker
than those of the model with $\rm A_{V,e}=3$ mag at 500,000 years.
This is caused by the dissociative recombination of HCO$^+$ with electrons 
at large radii in the model with the lower $\rm A_{V,e}$.

The evolution of the ortho-H$_2$CO 6 cm line is shown in Figure 20.
The absorption in that line occurs against the cosmic background radiation 
(CBR) because of a collisional pumping that lowers the excitation temperature 
below $T_{CBR}$ (Townes and Cheung 1969, Zhou et al. 1990).
The pumping works up to about $10^4$ cm$^{-3}$ in density, so the line goes
into emission at sufficiently high densities if the abundance of H$_2$CO is
constant through a entire core.
However, the abundance profiles of H$_2$CO are complicated with many variations.
As shown in Figure 9, H$_2$CO shows significant freeze-out before the dust
temperature reaches the CO evaporation temperature.
In addition to the variation of H$_2$CO with time, its abundance profile varies
with $\rm A_{V,e}$, that is, the cloud environment.
Therefore, the evolution of the line depends on the external visual excitation. 
In a lower $\rm A_{V,e}$ environment, the abundance drops sharply at outer 
radii, where the line cannot be thermalized (Figure 26). In that region, the 
pumping mechanism works efficiently due to the low abundance and the high 
kinetic temperature so that deeper absorptions are produced.
On the other hand, at higher $\rm A_{V,e}$, H$_2$CO is protected against the 
photodissociation, and the lower kinetic temperature reduces the collisional 
pumping to produce an emission line at 6 cm. 
Based on this result, we can qualitatively
distinguish the different environments of a star-forming
core; an emission line at higher $\rm A_{V,e}$ and an absorption 
line at lower $\rm A_{V,e}$.
We see only absorption lines toward the centers of three PPCs in K. Young et al.
(2004), which suggests that the cores are embedded in relatively low 
$\rm A_{V,e}$.
This is consistent with the higher $G_0$ that they found from the CO 
observations, compared to what is expected from the $\rm A_{V,e}$ found
in the dust radiative transfer calculation.

Figure 21 shows the evolution of the $\rm N_2H^+$ $1-0$ isolated hyperfine 
component ($1_{01}-0_{12}$).
The line has a hyperfine structure with 7-components, with 6 components 
located close to each other.
The $\rm N_2H^+$ abundance increases until CO evaporates from grain 
surfaces. Therefore, $\rm N_2H^+$ can be a good tracer of the density structure
in the earlier evolutionary stages before $T_D$ reaches $T_{evap}$ of CO.
Once CO starts to evaporate ($t=50,000$ years), the $\rm N_2H^+$ line becomes 
weaker.
$\rm N_2H^+$ is abundant even at radii larger than 0.03 pc in the model with 
$\rm A_{V,e}=3$ mag (see Figure 12), so the line becomes optically thick and 
shows the self-absorption profiles even in earlier time steps. 
However, the blue asymmetry in this line traces the infall at the radii outside
the CO evaporation front because the $\rm N_2H^+$ abundance
drops by a few orders of magnitude inside the CO evaporation front. 

\section{DISCUSSION}
In actual observations, we have to choose a proper beam size, transition, and 
observational technique such as central position observation or mapping
in order to get correct information about star formation.  
HCN could be a good probe for distinguishing Class 0 from the FHSC, 
as described above.
However, we have to use a very good resolution for the classification.
Figure 22 shows the difference between simulated observations of the HCN $1-0$ 
line with two beam sizes of 5$\as$ and 50$\as$. 
The HCN $1-0$ has a hyperfine structure of 3-components, which are well 
separated from each other. 
In the observation with a lower resolution, it is difficult to tell when the 
transition between the FHSC and Class 0 occurs.
On the other hand, a big change in the line strength between two stages is seen 
in the simulation with a high resolution.
Therefore, an interferometer observation is needed to differentiate Class 0 from
the FHSC with HCN lines. 

The chemical evolution is affected by the environment of cores as shown in
Figure 12 and 13, as well as grain properties inside cores as shown in Figure 
14 and 15.
The environment of star-forming cores can be described by the combination
of the external visual extinction and the strength of the UV field.
The UV field is extincted differently by different grain properties along 
lines of sight, such as the material and size of dust grains, and by 
different cloud geometries. 
$\rm A_{V,e}$ is important in the chemistry, and the strength of the
UV field is important in the gas energetics.

In the following we compare how various environments affect molecular line 
profiles. 
Figure 23 shows the comparison of the C$^{18}$O $1-0$, $2-1$, and $3-2$ lines
in the models of two different visual external extinctions ($\rm A_{V,e}=0.5$ 
and 3 mag) at 200,000 years.
Here, $G_0$ is calculated with the equation shown in \S 2.3 assuming
the same grain properties and cloud geometry.
All lines are simulated toward the center of the model core.
We use the same beam size for each line as those of J{\o}rgensen et al. (2002);
33$\as$, 21$\as$, and 14$\as$ for $1-0$, $2-1$, and $3-2$, respectively.
The $1-0$ line has similar strength to the $2-1$ line, which agrees with the 
actual observations and the simulated lines with an abundance profile of 
a drop function (J{\o}rgensen et al. 2004b).     
The wing structure in the C$^{18}$O lines is due to the infall velocity
structure rather than an optical depth effect.
The optical depths of the lines are $\ll 1$.
The $1-0$ line does not change with different $\rm A_{V,e}$, but the
$2-1$ and $3-2$ lines are weaker for higher $\rm A_{V,e}$.
Therefore, multi-transition observations can be good probes of the cloud 
environment.
 
Figure 24 compares the ortho-H$_2$CO 6 cm lines along various lines of sight 
at 50,000 years to show the effect of different $\rm A_{V,e}$.
For smaller $\rm A_{V,e}$, H$_2$CO has a sharp drop in the abundance profile due
to the photodissociation, and $T_K$ is high because of the 
PE heating at outer radii as shown Figure 26.
Therefore, the excitation temperature of the 6 cm line for the model with the
smaller $\rm A_{V,e}$ is much lower than the temperature of the CBR 
($T_{CBR}=2.7$ K) at larger radii than 0.025 pc.
The 6 cm line produces deep absorption at small impact parameters, but
the absorption dip is much weaker at a large impact parameter because of the low
abundance at large radii.   
On the other hand, in the core with greater $\rm A_{V,e}$, even though the 
H$_2$CO abundance is very high at outer radii, the absorption dip is not deeper 
than at inner radii because the excitation temperature is similar to 2.7 K.
As a result, the 6 cm line goes into emission at the center.   

We show the effect of different UV strengths in Figure 25.
We use the same $\rm A_{V,e}$ of 3 mag, but two different $G_0$ have been 
considered.
One is obtained from the equation in \S 2.3 ($G_0=0.005$), and the
other is assumed to be 10 times bigger ($G_0=0.05$).  
The abundance profiles of two models are almost the same because the chemistry 
is not affected much by the kinetic temperature at later time steps as shown 
in Figure 26. 
However, the kinetic temperature of the model with $G_0=0.05$ is higher than 
that of the model with $G_0=0.005$ at outer radii.
In the model of $G_0=0.05$, the excitation temperature of the 6 cm line
is below $T_{CBR}$ at larger radii than 0.04 pc. 
Therefore, the high abundance and the lower excitation temperature than 2.7 K
at outer radii produce deep absorption lines at all impact parameters.
These comparisons suggest that we need to do multi-position observations 
as well as multi-transition observations in order to test the cloud environment.

We have to note that there is a caveat about collision rates. Collision rates,
in theory, have been calculated down to 10 K, but, in our model, 
$T_K$ goes down to about 6 K at the center in the PPC stage and the FHSC stage.
We, therefore, extrapolate the collision rates linearly to 5 K. 
We also emphasize that outflows and hot cores, where the dust temperature is 
higher than 100 K, are not considered in this work, 
and they can affect the interpretation of lines from the protostellar stage. 

\section{CONCLUSIONS}
Although we still have limitations such as 1-D geometry and no grain 
surface chemistry in this proposed model, we certainly see that 
the chemical timescale
is not short or long enough to ignore the abundance variations of species 
during the dynamical evolution in star formation, especially in the inner dense 
regions as shown in Table 4.
The abundances of molecules vary by a few orders of magnitude with time at
the central part, but the variation becomes less significant in outer regions.
Therefore, abundance profiles at given times are very wiggly.
Even though there are small bumps and wiggles in the abundance profiles of
molecules, the overall structure of the profiles can be understood as
interactions between freeze-out and evaporation of the molecules, themselves,
and CO.
Some of the bumps, jumps, or drops are due to CO evaporation. The
presence or absence of CO from the gas phase has a tremendous effect on the
less abundant species.

Astronomers use molecular lines to study star-forming regions.
However, the line observations can be misunderstood if one does not consider
the abundance variations with time or with radius, which have been shown in
this study.
According to our results, carbon- or sulfur-bearing molecules such as CO, CS,
and H$_2$CO  readily freeze-out from the gas
on all kinds of dust grains
as shown in other theoretical modeling work.
HCO$^+$, a daughter species of CO, also disappears from the gas phase due to
the CO freeze-out.
However, nitrogen-bearing molecules such as NH$_3$ and $\rm N_2H^+$ could be
a good probe of the density structure in the PPC stage if grain mantles are
covered with SiO$_2$ or CO dominantly. On the other hand, they are 
frozen-out/depleted significantly on water dominant grain mantles at very late 
times of the PPC stage or at the FHSC stage.
HCN is an interesting species. It exhibits strong freeze-out on silicate or
water dominant mantles as other carbon-bearing species, but it becomes abundant
with time on CO dominant grain mantles at $t<0$.
Therefore, we may use HCN with NH$_3$ or $\rm N_2H^+$ to learn about grain
properties in PPCs.
In addition, HCN becomes abundant very quickly in the transition from the FHSC
to Class 0, so it could be a probe to distinguish the two evolutionary stages.
In the later stages than the FHSC, the assumption of a constant abundance or
a simple functional form of the abundance structure in some molecules can lead
one to incorrect density and velocity structures because abundance profiles of
those molecules vary significantly with time and radius, so they are very
wiggly, especially, in the very inner region.
We, therefore, suggest that the chemical evolution should be considered
simultaneously with the dynamical evolution when one studies the evolution in
star formation, and high spatial resolution maps are very important.
However, one can still adopt simplified functional forms of abundance profiles
depending on the types of molecules. Molecules such as CO, HCO$^+$, NH$_3$,
and $\rm N_2H^+$ show rather simple abundance structures that can be
approximated by a step function or a drop function.

In our study we find strong evidence for line wings produced
in the infalling envelopes. The primary reason for the presence of the infall
wings is the sharp rise of molecular abundances within the specific
evaporation front for that species.  However, these same species can likely
be liberated from grains within the outflow, which can hide the presence
of wings produced solely from infall.

The complicated abundance profiles of HCN, CS, and H$_2$CO
highlight another important result of this work. To first-order, abundance
profiles of species in star-forming cores can be approximated using simple
three component profiles (high in center, freeze-out in middle, and
an abundance rise in outer layers).
However, there are important second-order effects where
abundances in the gas are significantly altered by the evaporation of separate
species.
These second-order effects could be important if, for example, they resonate
with particular changes in the velocity field and may eventually account for
some difficulties in the interpretation of observations.
The influence of CO on the chemistry of CS is one example to show how the
second-order effect is coupled with the velocity field in collapse as
shown in \S 4.
J{\o}rgensen et al. (2004b) calculated correlations between various molecular
abundances and envelope masses of star-forming cores.
According to their analysis, CO and
HCO$^+$ show very high correlation with the envelope mass
(i.e. evolutionary stage), but SO and HCN show little correlation with the
envelope mass.
These results could be explained by the second-order effects seen in our models.
CO does not have the second-order effect, and the abundance of HCO$^+$ is
directly linked to CO.
Therefore, CO and HCO$^+$ both become more abundant because of the CO
evaporation as the envelope mass decreases.
However, HCN is affected by CO and CN evaporation, and the second-order effects
may complicate the relation between the HCN abundance and the envelope
mass.
For the SO chemistry, oxygen plays a role. Oxygen is locked onto
grain surfaces in the form of water, and sulfur-bearing species, including SO,
react with carbon, which is produced by the evaporated CO and He$^+$ ions,
to form CS.
As a result, the evaporation of SO does not result in the high abundance of SO
in the gas phase, so a tight correlation between SO and the envelope mass is not
seen.

Different timescales during the PPC stage and different dynamics during collapse
need to be explored in future works.
In this work, a few simple tests of different timescales, that is, different
density evolutions, during the PPC stage have been done.
According to the results, the freeze-out/depletion dip is shallower and
narrower during
the PPC stage if the timescale of the PPC stage is shorter than that of
standard model, and vice versa.
However, during collapse, the depth of freeze-out/depletion dip is the same
regardless of the timescale of the PPC stage, and the difference between the
various timescales becomes small with time.
CS and HCN are the most sensitive molecules to the density evolution during the
PPC stage. Their abundances vary more than one order of magnitude depending on 
the timescale of the PPC stage.

As separate work, we are also comparing this model with actual 
observations. 
We have compared the result of H$_2$CO with the ortho-H$_2$CO 6 cm line 
toward three different PPCs (Young el al. 2004).
Evans et al. (2004) studies the chemical status of B335 by comparing
multi-transition observations of various molecules with simulated ones with
chemical models of different dust properties and initial conditions.
Comparisons of the results of this work with Multi-transition and 
multi-position observations done with single dishes and interferometers 
in more Class 0 and I sources will appear in separate papers (Lee et al.
2004a, 2004b). 

\section{Acknowledgments}
We are grateful to Chad Young for providing his code for the luminosity 
evolution in a star-forming core.
We are also grateful to Steve Doty for letting us use his gas energetics 
code.
We thank E. van Dishoeck for many helpful comments.  
We thank the NSF (grants AST-9988230, 0307250, 0335207) for support.

\clearpage

\begin{deluxetable}{lccccc}
\tablecolumns{6}
\footnotesize
\tablecaption{\bf The summary of the standard model
\label{tab1}}
\tablewidth{0pt}
\tablehead{
\colhead{$\rm R_i$\tablenotemark{a}}                &
\colhead{$\rm R_o$\tablenotemark{b}}    &
\colhead{a\tablenotemark{c}}              &
\colhead{$\rm M_{core}$\tablenotemark{d}} & 
\colhead{$\rm A_{V,e}$\tablenotemark{e}} & 
\colhead{$G_0$\tablenotemark{f}}              \\
\colhead{pc}                      &
\colhead{pc}        &
\colhead{km s$^{-1}$}                  &
\colhead{\msun} &
\colhead{mag} &
\colhead{} 
}

\startdata
0.00033& 0.15 & 0.22 & 3.6 & 0.5 & 0.4
\enddata

\tablenotetext{a}{Inner radius of the core}
\tablenotetext{b}{Outer radius of the core}
\tablenotetext{c}{Sound speed for calculating density structures in inside-out
collapse}
\tablenotetext{d}{The initial mass of core}
\tablenotetext{e}{External Visual extinction}
\tablenotetext{f}{The strength of the UV field relative to the local ISRF}

\end{deluxetable}

\begin{deluxetable}{lc}
\tablecolumns{2}
\footnotesize
\tablecaption{\bf The surface binding energy of species onto SiO$_2$ dust grain mantle  
\label{tab2}}
\tablewidth{0pt}
\tablehead{
\colhead{species}        &
\colhead{binding energy (K) } 
}
\startdata
H$_2$O & 5700\\
CS & 1999  \\
H$_2$CO & 1722 \\
HCN & 1722 \\
CN & 1476 \\
CO & 1181 \\
NH$_3$ & 1082 
\enddata
\end{deluxetable}

\begin{deluxetable}{lc}
\tablecolumns{7}
\footnotesize
\tablecaption{\bf Initial elemental abundances
\label{tab3}}
\tablewidth{0pt}
\tablehead{
\colhead{}                &
\colhead{Abundance}              \\
\colhead{Element}                  &
\colhead{(Relative to H$_2$)}
}
\startdata
He & 0.28 \\
He$^+$ & $1.33\times 10^{-10}$ \\
C$^+$ & $1.46\times 10^{-4}$ \\
N & $4.28\times 10^{-5}$ \\
O & $3.52\times 10^{-4}$ \\
Si$^+$ & $4.0\times 10^{-11}$ \\
Mg$^+$ & $6.0\times 10^{-11}$ \\
S$^+$ & $4.0\times 10^{-8}$ \\
Fe$^+$ & $6.0\times 10^{-11}$ \\
Na$^+$ & $4.0\times 10^{-11}$ \\
P$^+$ & $6.0\times 10^{-9}$ 
\enddata
\end{deluxetable}

\begin{deluxetable}{lccc}
\tablecolumns{4}
\footnotesize
\tablecaption{\bf Characteristic Timescales
\label{tab4}}
\tablewidth{0pt}
\tablehead{
\colhead{}                &
\colhead{Process}         &
\colhead{Timescale (yrs)} &
\colhead{Timescale (yrs)} \\
\colhead{}                &
\colhead{}                &
\colhead{}                &
\colhead{for $n(\rm H_2)=10^5$ cm$^{-3}$}
}
\startdata
Dynamics & Bonnor-Ebert spheres (t$<$0) \tablenotemark{a}& 
dependent on central densities & $6\times 10^4$ \tablenotemark{b}\\
 & Free-fall (t$>$0) & $\frac{2.4\times 10^7}{\sqrt {n(\rm H_2)}}$ & 
$7.6\times 10^4$ \\
\hline
Dust energetics & Radiative heating & $r/c$ \tablenotemark{c} & 0.48 \\ 
\hline
Gas energetics & Gas-dust collision & $\frac{1.26\times 10^8}{\sqrt{T_K} 
n(\rm H_2)^2}$ & 0.004 \tablenotemark{d}\\
\hline
Chemistry & Gas-phase chemistry \tablenotemark{e} & 
$\frac{2.6\times 10^9}{n(\rm H_2)}$ & $2.6\times 10^4$ \\ 
 & Freeze-out & $\frac{6\times 10^9}{n(\rm H_2)} \sqrt{\frac{m}{T_K}}$ 
\tablenotemark{f} & $10^5$ \\ 
 & Evaporation (desorption) & $\frac{m}{1.6\times 10^{20}} 
exp(\frac{E_b}{kT_D})$ \tablenotemark{g}
& $6.5\times 10^{31}$ ($T_D = 10$ K) \tablenotemark{h} \\
 & & & 0.004 ($T_D=30$ K)\\
\hline
Excitation of molecular & Collisions & $\sim \frac{3.2\times 10^3}{\sqrt{T_K} 
n(\rm H_2)}$ & $0.013$ \tablenotemark{i} \\   
energy levels & & &
\enddata
\tablenotetext{a}{ 
For $t<0$, we have assumed a sequence of Bonnor-Ebert spheres with a total
time fixed at $10^6$ years. As the central density of the Bonnor-Ebert sphere
increases, the timescale decreases.    
}
\tablenotetext{b}{We adopt the 
timescale for the Bonnor-Ebert sphere of $n_c=10^5$ cm$^{-3}$. 
}
\tablenotetext{c}{$r$ ($=0.15$ pc) is the size of our model core, and $c$ is 
the speed of light.}
\tablenotetext{d}{$T_K=10$ K is used for all timescales.}
\tablenotetext{e}{Induced by the cosmic-ray ionization}
\tablenotetext{f}{$m$ is the molecular weight.}
\tablenotetext{g}{$E_b$ is the binding energy of a molecule onto the 
surface of dust grains.}
\tablenotetext{h}{For CO}
\tablenotetext{i}{De-excitation timescale of the transition of CO J$=1-0$. 
The de-excitation rate is adopted from Flower and Launay (1985).} 
\end{deluxetable}

\begin{figure}
\figurenum{1}
\plotone{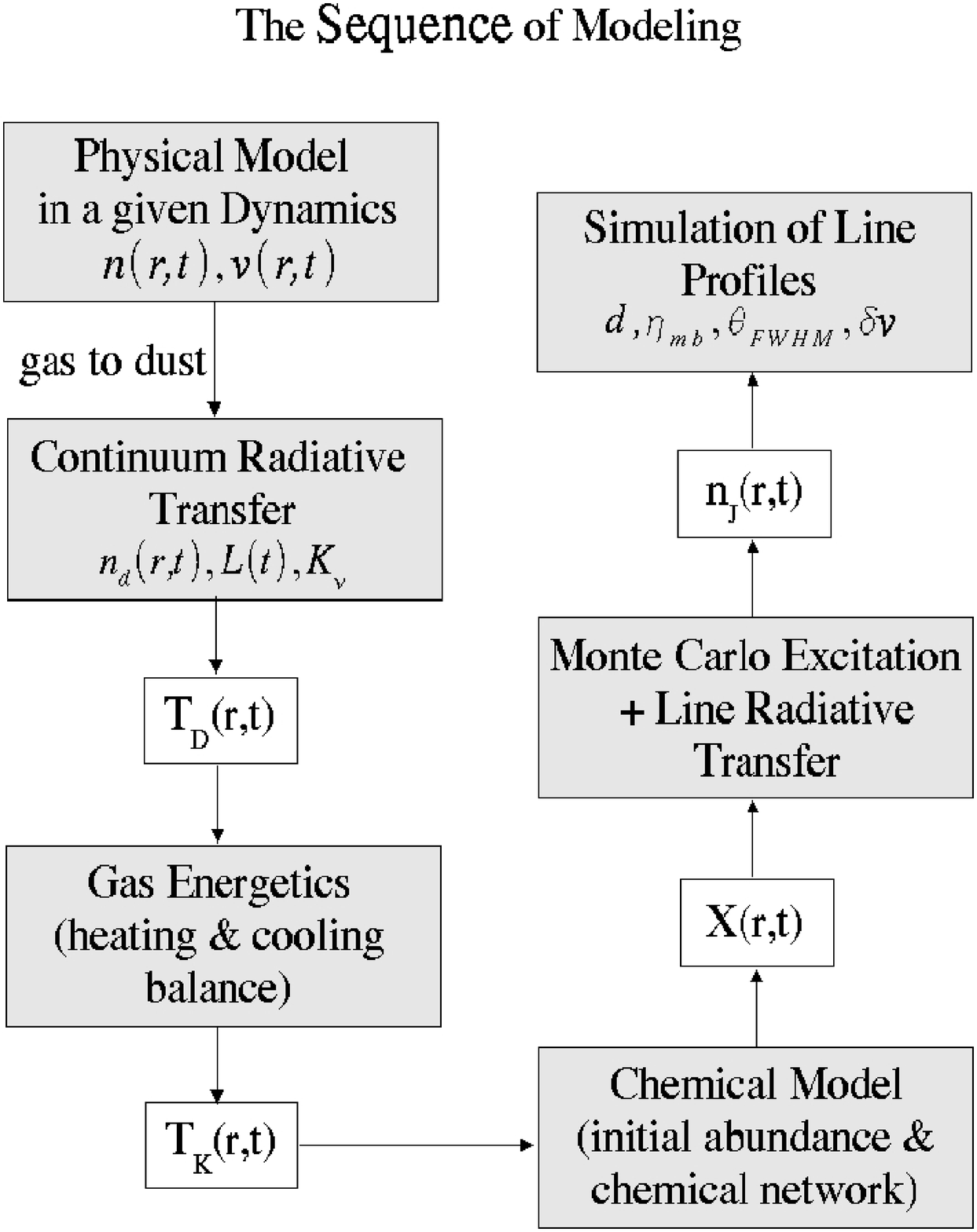}
\figcaption{
The flowchart to show the sequence of the whole modeling process.
}
\end{figure}

\begin{figure}
\figurenum{2}
\plotone{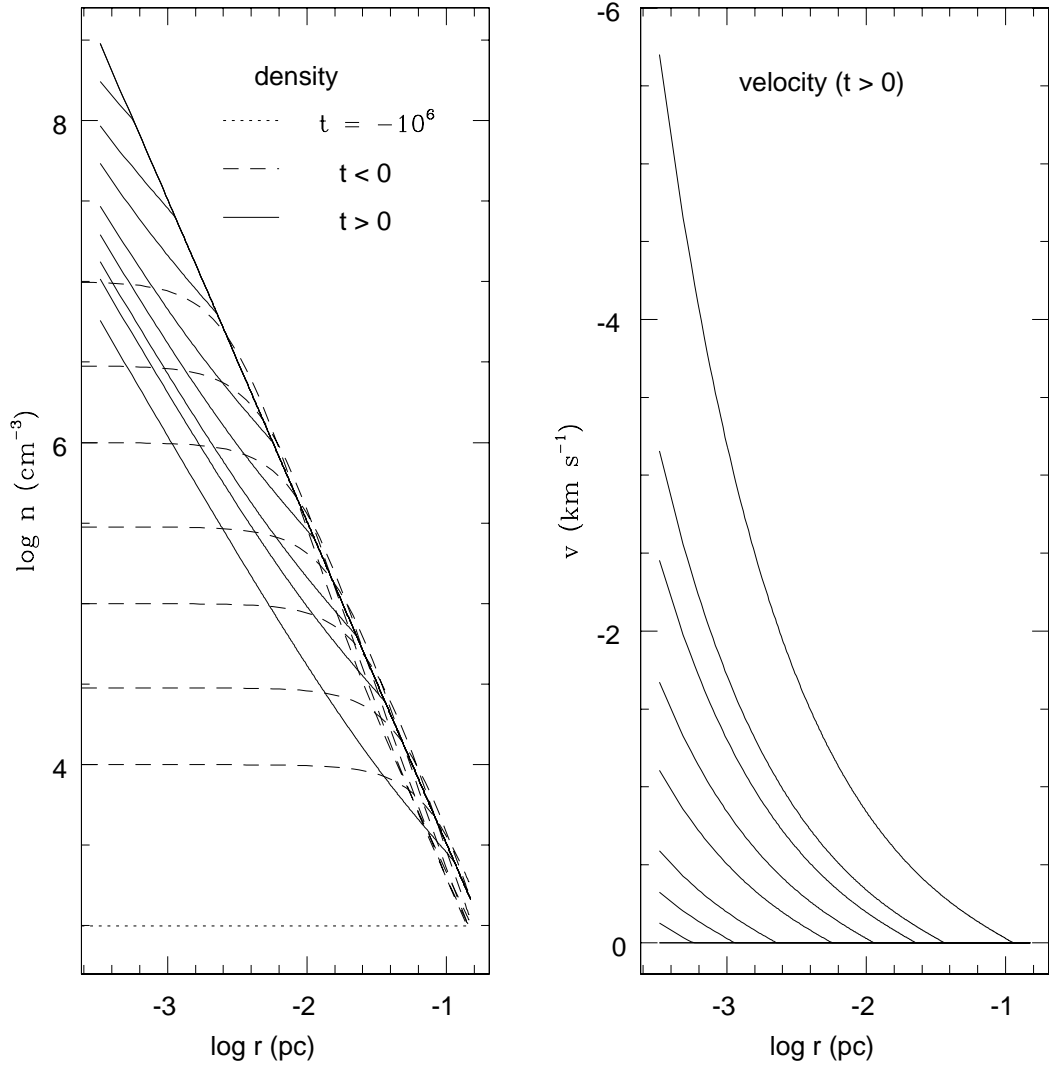}
\figcaption{
Density and velocity profiles of the model core with time. In the left box, 
dashed lines represent density profiles of Bonnor-Ebert spheres of $n_c=10^4$,
$3\times10^4$, $10^5$, $3\times10^5$, $10^6$, $3\times10^6$, and $10^7$.
The dotted line indicates the very initial constant density profile.
Solid lines represent density profiles during collapse at 0,  
$2.5\times10^3$, $5\times10^3$, $10^4$, $2.5\times10^4$,
$5\times10^4$, $10^5$, $1.6\times10^5$, and $5\times10^5$; 
the timescale during collapse is divided into 190 time steps.
The right box shows velocity profiles of collapse at the same time
steps as in density profiles.
}
\end{figure}

\begin{figure}
\figurenum{3}
\plotone{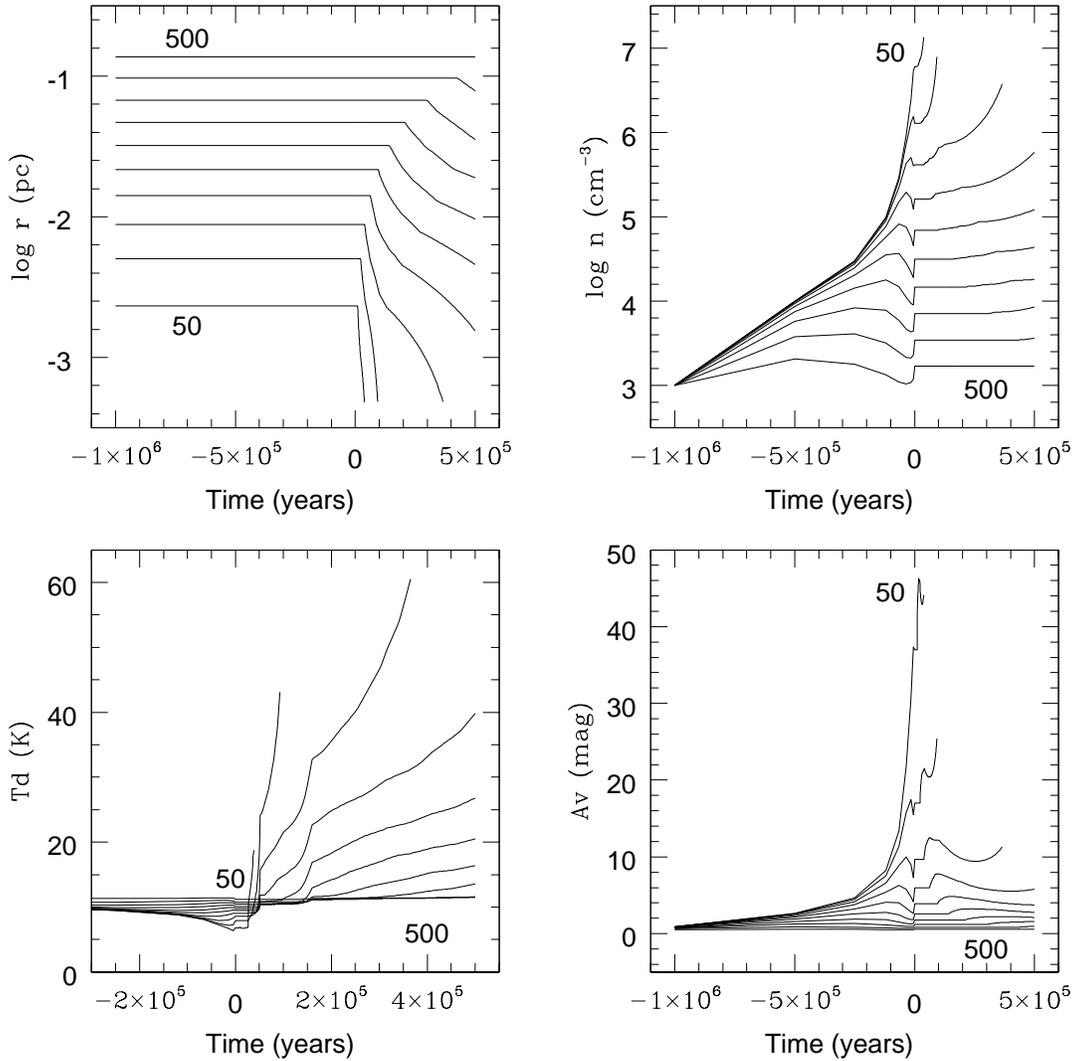}
\figcaption{
\small Physical parameters of given parcels of gas with time; the left upper 
box for
positions, the right upper box for densities, the left lower box for dust
temperatures, and the right lower box for total visual extinctions, 
respectively.
Each line represents the parcel of the grid number of 50, 100, 150, 200, 250,
300, 350, 400, 450, and 500 among 512 grid points.  
A time of zero indicates the starting point of the gravitational collapse of the
model core. In the position plot, a parcel stays in a given position until the
infall wave reaches its initial position, and then moves inward with time.
There is a discontinuity in density and $\rm A_V$ at $t=0$ because we 
combine the Bonnor-Ebert spheres and inside-out collapse artificially. 
In the dust temperature plot, before the beginning of collapse, the 
temperature of a given parcel decreases with time because density and visual 
extinction increase, so the ISRF, the only heating source of PPCs, cannot 
penetrate well.
However, after collapse starts, the core is heated by accretion, 
and temperatures go up quickly to produce a big jump around 50,000 years.
There is another sharp increase of temperature caused from the turn-on of
photospheric luminosity from gravitational contraction and deuterium burning
in the central star at 100,000 years (see \S 2.2).
X-axis of the dust temperature plot is not the same as that of other plots.
The variation of dust temperature is small in early stages, so we plot it from
t=-300,000 years to show better resolution in later stages.  
}
\end{figure}

\begin{figure}
\figurenum{4}
\plotone{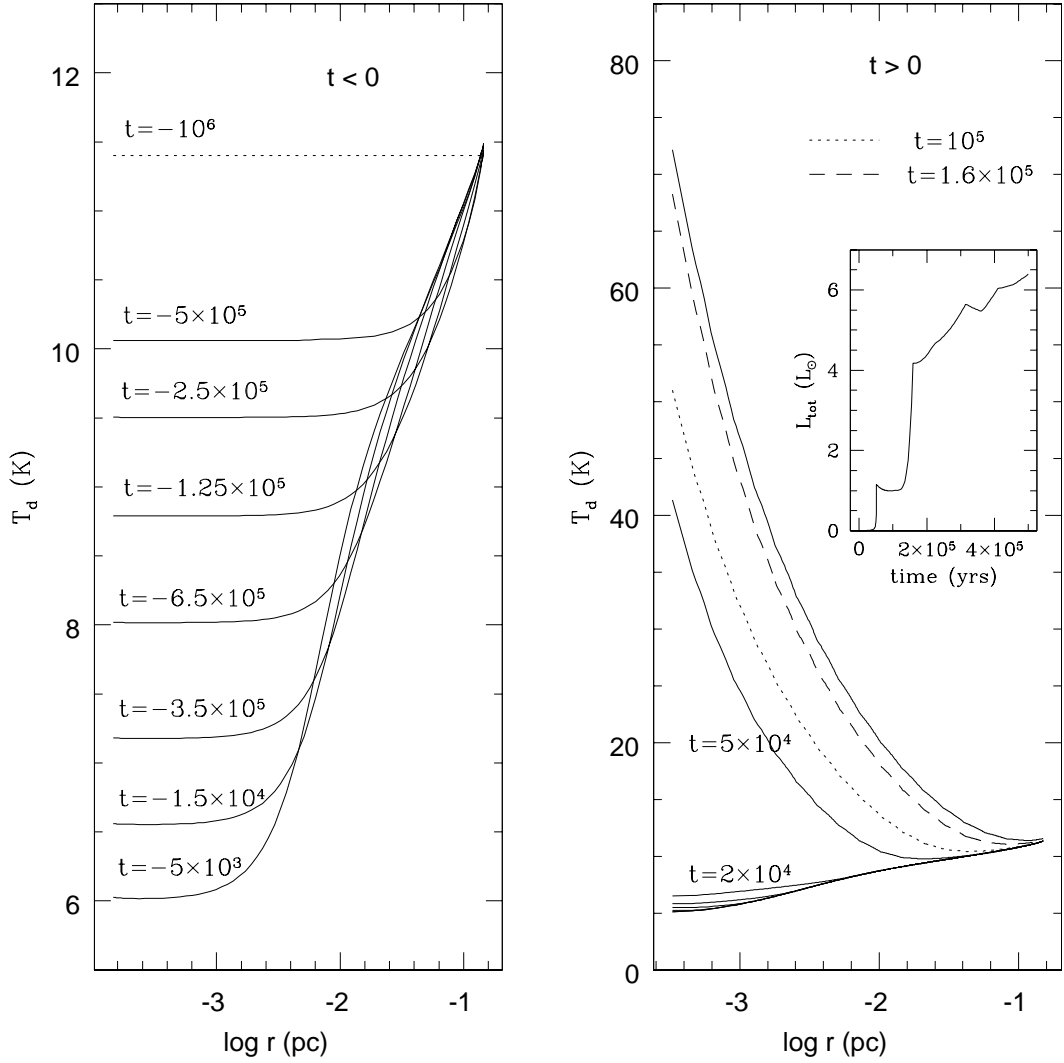}
\figcaption{
The results of dust continuum radiative transfer calculation. The left box shows
dust temperature profiles of Bonnor-Ebert spheres of seven different central
densities same as in Figure 2. The dotted line in this box represents the 
temperature profile for the initial density profile.
In the right box, lines represent dust temperature profiles during
collapse at the same time steps as in Figure 2. 
There is a big jump between 20,000 and 50,000 years. The model core is 
in the stage of the first hydrostatic core until 20,000 years.
At that stage, accretion luminosity is still small because of a large radius 
(5 AU) of the central source; consequently the dust temperature stays very low. 
The dotted line and dashed line show another jump in temperature caused from
the turn-on of photospheric luminosity at 100,000 years.
The small box inside the right box shows the evolution of the total luminosity,
which is the sum of the accretion luminosity, the disk luminosity, and the 
protostellar luminosity, in the unit of the solar luminosity.
}
\end{figure}

\clearpage

\begin{figure}
\figurenum{5}
\plotone{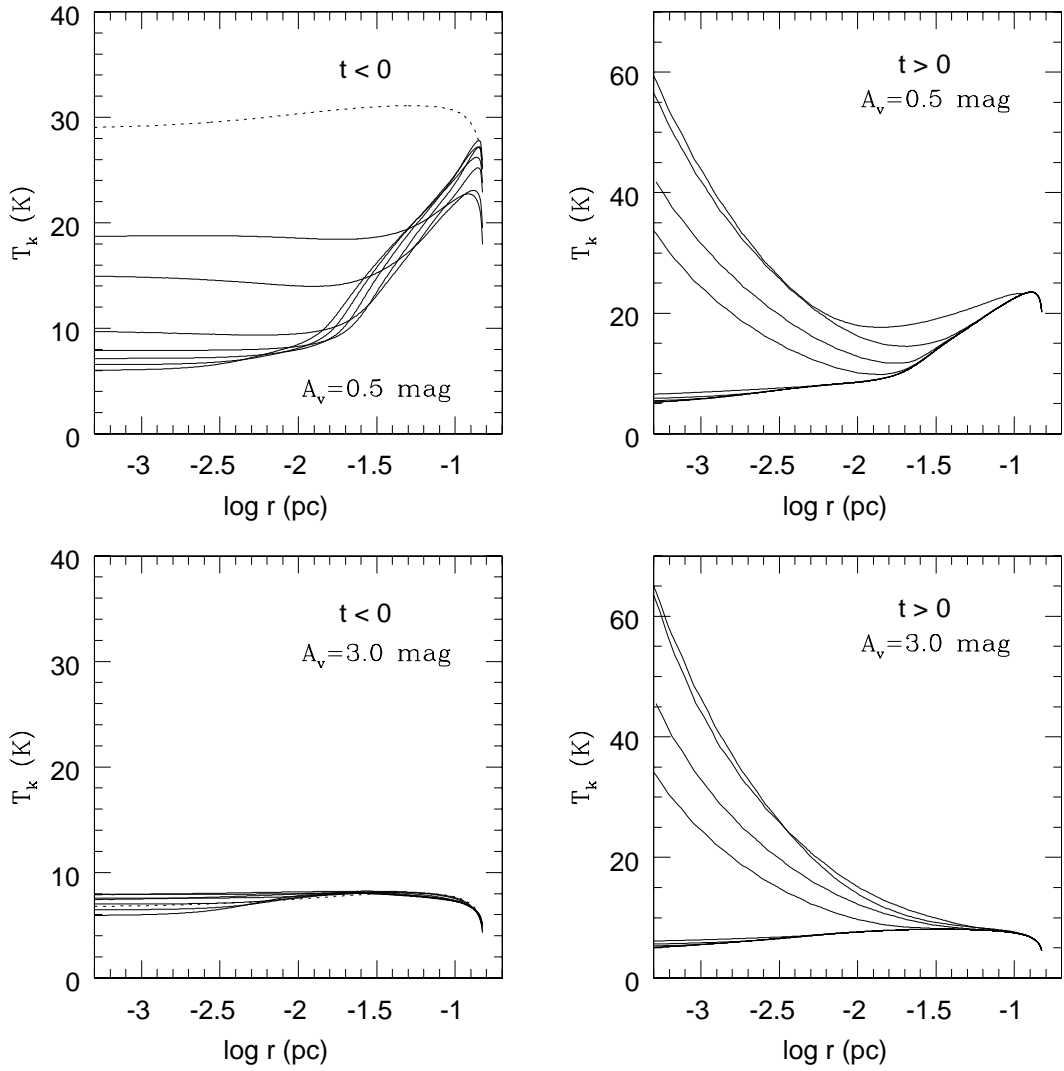}
\figcaption{Kinetic temperatures calculated with the energetics code. 
The selected time steps are the same as in Figure 2 and 4.
Left two boxes show gas temperatures in seven Bonnor-Ebert spheres. 
The dotted lines
indicate the gas temperatures calculated in the initial state. 
Gas temperatures during collapse are represented in right boxes. 
The upper boxes and the lower boxes compare results of heating by the ISRF 
attenuated by different magnitudes of external visual extinction.
As the central density of the
B.E. sphere grows, the gas temperature at small radii gets close to the dust
temperature.  
}
\end{figure}

\clearpage
\begin{figure}
\figurenum{6}
\plotone{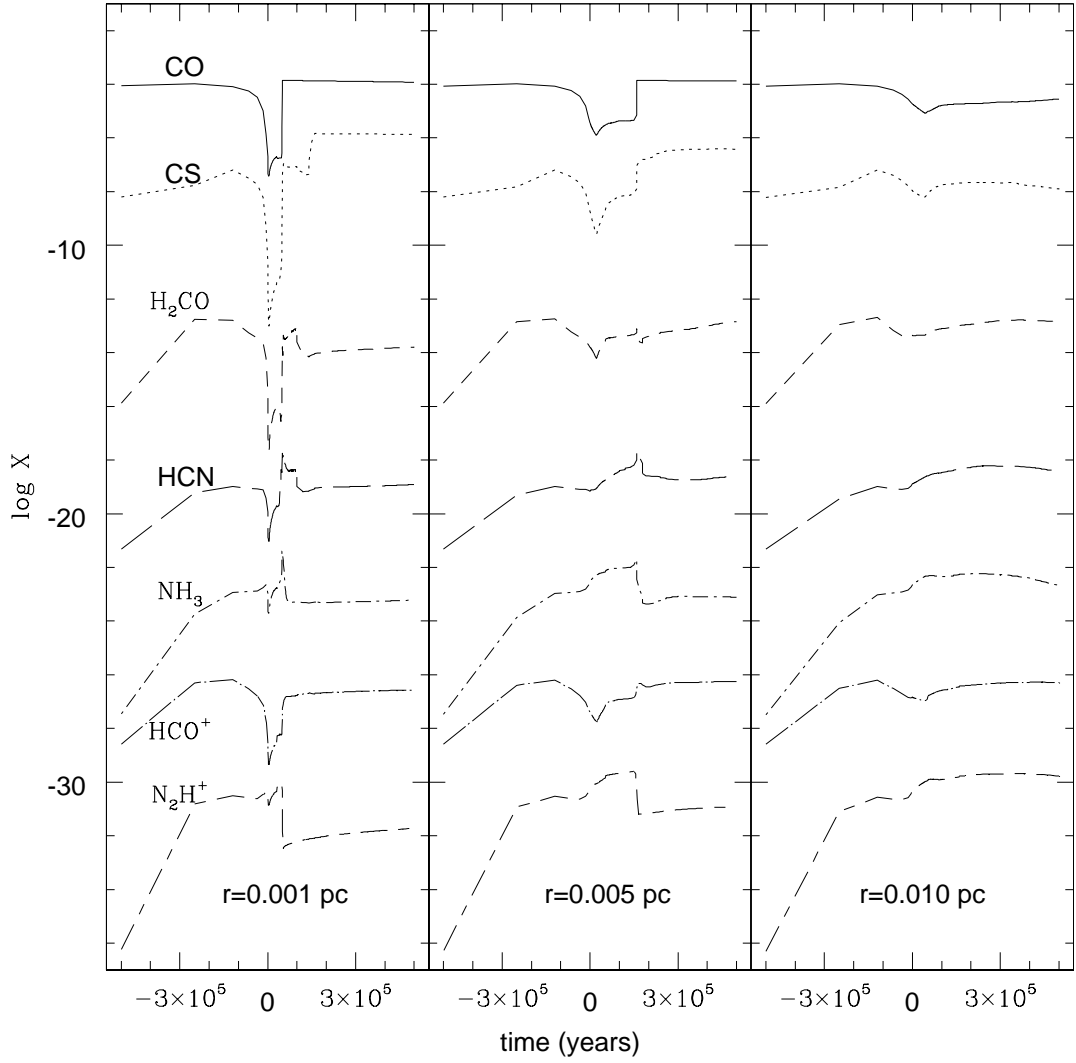}
\figcaption{ The evolution of abundances at given radii, r$=0.001$ pc,
0.005 pc, and 0.01 pc.
The lines are for CO (solid), CS (dot), $\rm H_2CO$ (short dash), HCN (long
dash), $\rm NH_3$ (dot - short dash), HCO$^+$ (dot -long dash), and
$\rm N_2H^+$ (short dash - long dash) from the top to the bottom.
Lines are shifted up and down to compare with CO evolution.
CS is shifted up by the 1.7 orders of magnitude, and HCN, NH$_3$, HCO$^+$, and 
$\rm N_2H^+$ are shifted down by the 5, 10, 15, 18, and 21 orders of magnitude,
respectively. 
}
\end{figure}

\clearpage

\begin{figure}
\figurenum{7}
\plotone{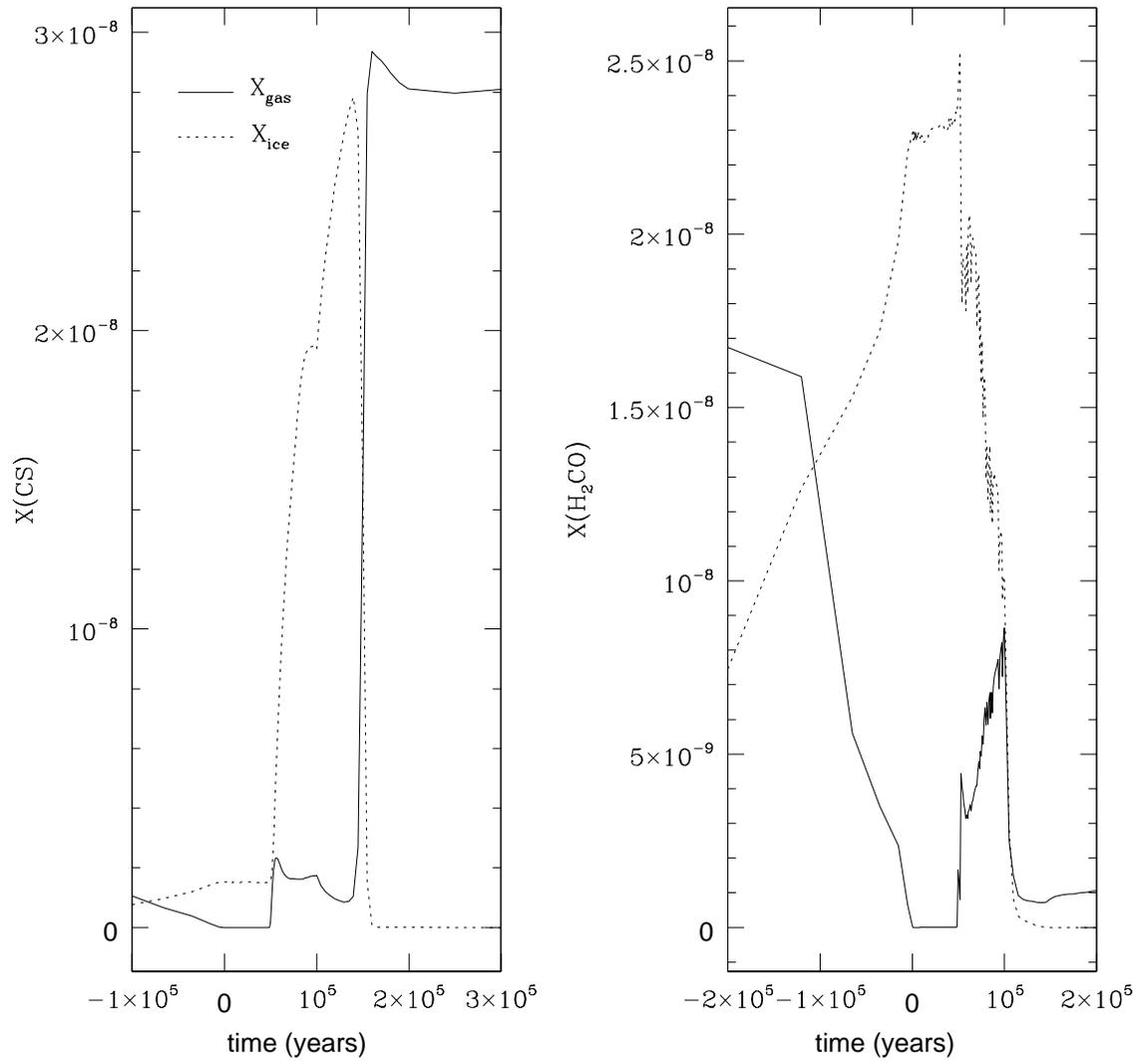}
\figcaption{The evolution of the abundance of CS and H$_2$CO at a given radius,
r$=0.001$ pc.
Solid and dotted lines represent abundances in gas and in ice versus time.
The y-axis is not a logarithmic scale, unlike in Figure 10.
}
\end{figure}

\begin{figure}
\figurenum{8}
\centering
 \vspace*{7.8cm}
   \leavevmode
   \includegraphics{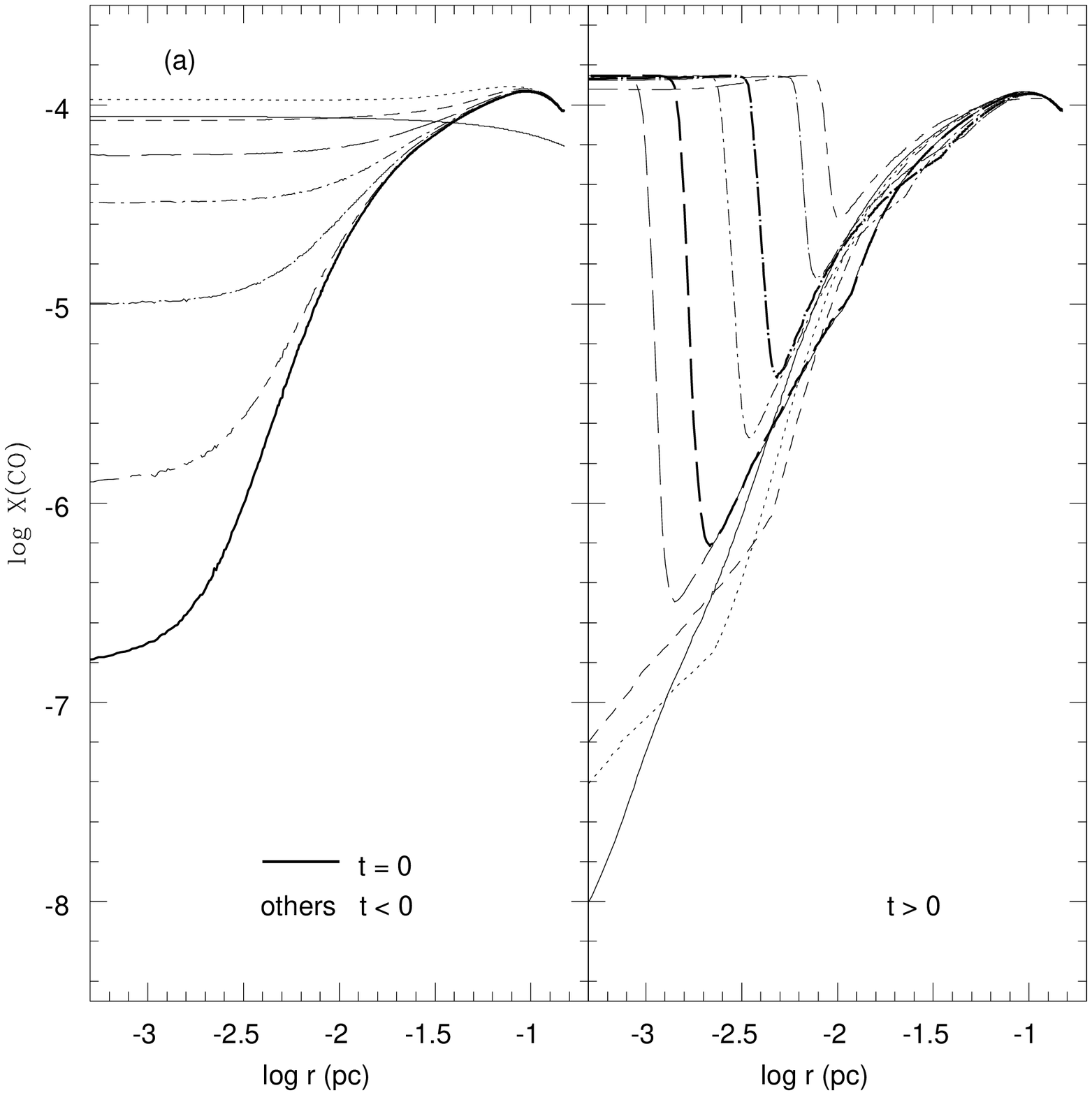}
   \includegraphics{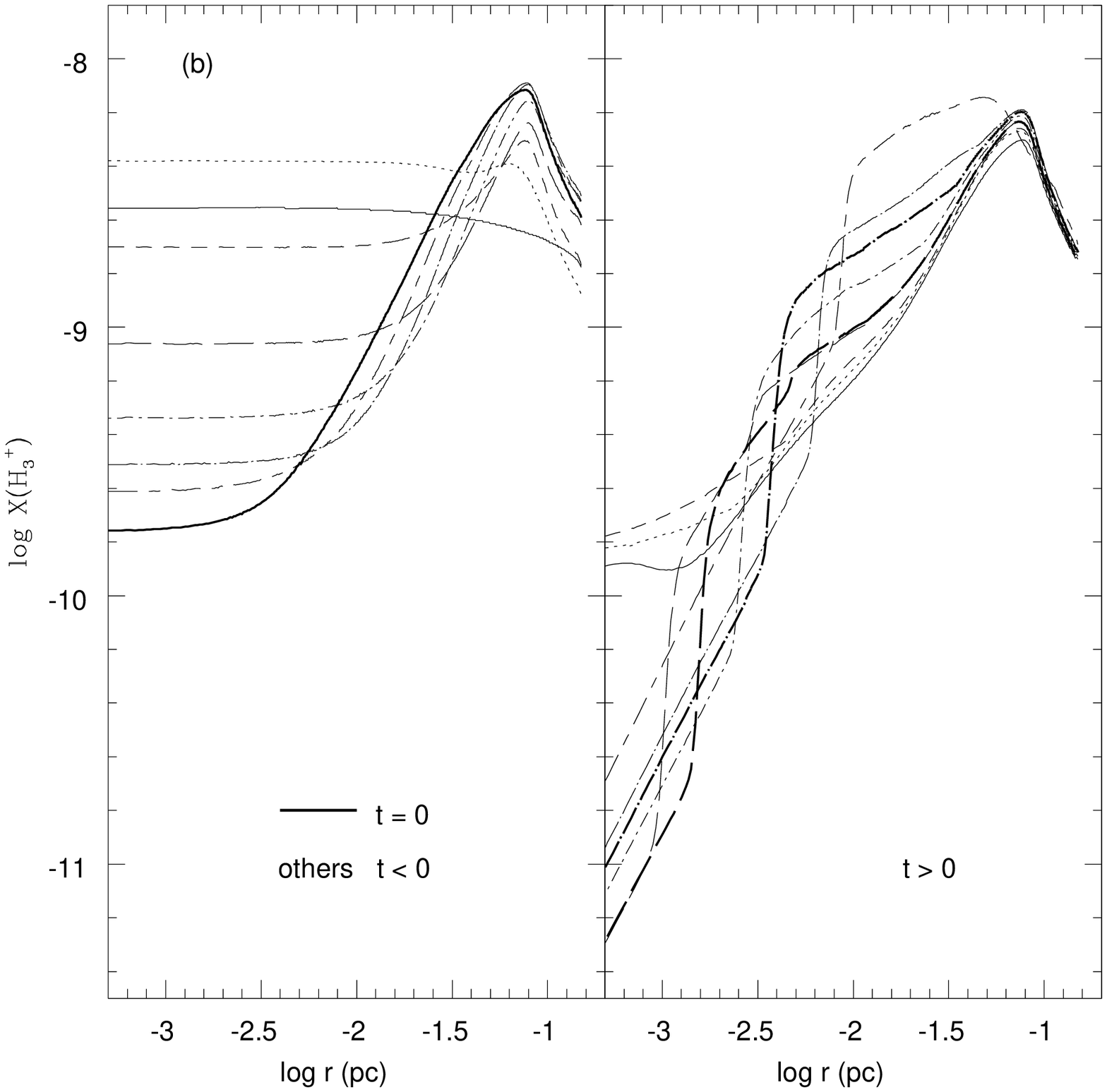}
   \includegraphics{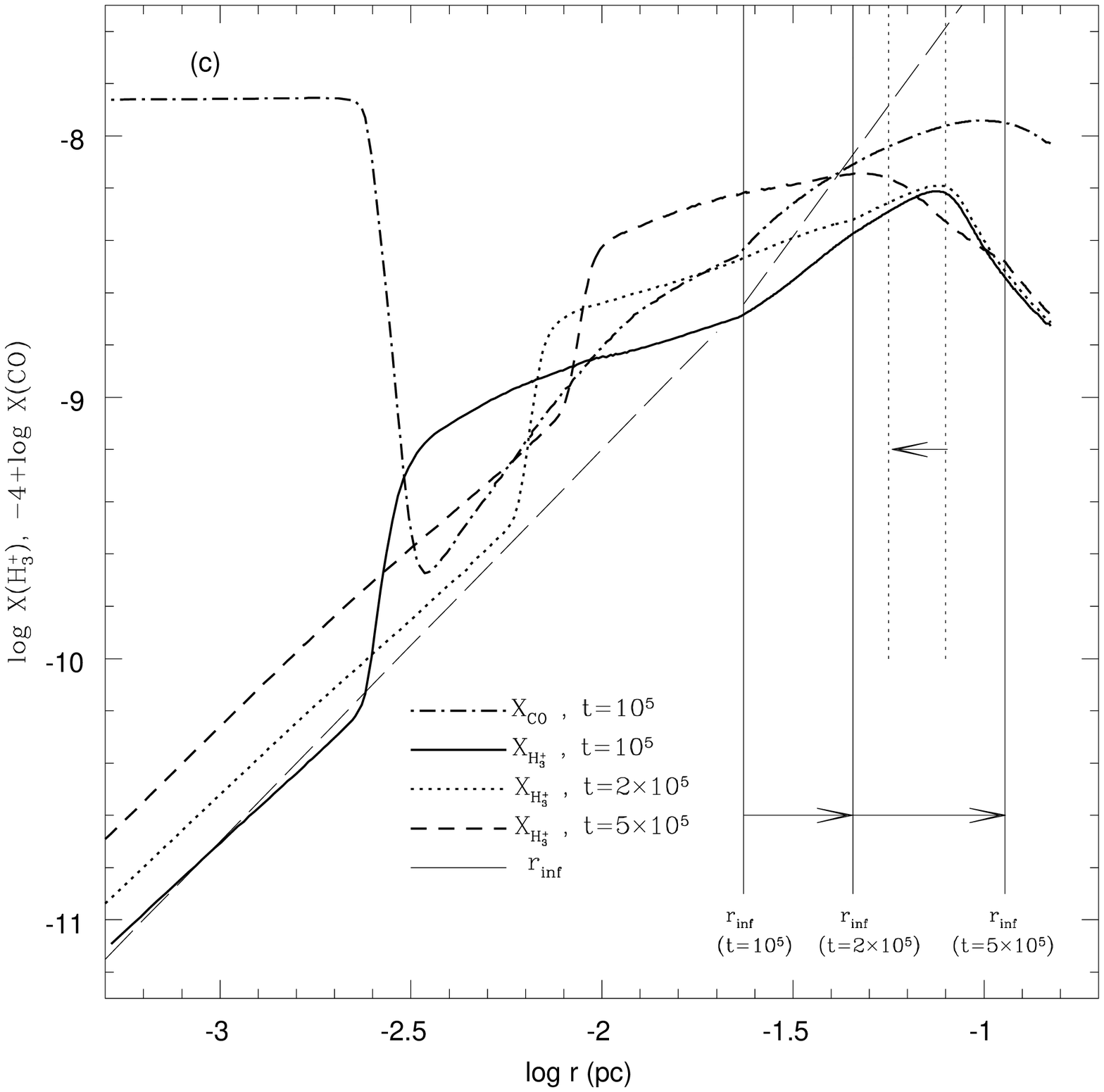}
\vskip 4.50in
\figcaption{
(a) and (b). CO and H$_3^+$ abundance profiles with time in some selected time steps; 
the left box for the evolution of Bonnor-Ebert 
spheres and the right box for the evolution during collapse in each window.
In the left box, lines are for time steps of $5\times10^5$ (thin solid),
$2.5\times10^5$ (dot), $1.25\times10^5$ (short dash), $6.5\times10^4$ 
(long dash), $3.5\times10^4$ (dot - short dash), $1.5\times10^4$ 
(dot - long dash), $5\times10^3$ (short dash - long dash) years 
before collapse, and the starting point of collapse (thick solid). 
The right box represents the CO abundance profiles at $10^3$ (solid), 
$10^4$ (dot), $2\times10^4$ (short dash), $5\times10^4$ (thin long dash), 
$5.2\times10^4$ (thick long dash), $10^5$ (thin dot - short dash), 
$1.5\times10^5$ (thick dot - short dash), $2.0\times10^5$ (dot - long dash), 
and $5\times10^6$ (short dash - long dash) years after collapse begins. 
(c). The effect of CO and dynamical evolution to the H$_3^+$ abundance profiles.
The long dashed line indicates the H$_3^+$ abundance profile if the H$_3^+$
density is constant. The thin dotted lines show how the abundance profile
established in an earlier time step affects an abundance profile in a later
time step.
}
\end{figure}

\begin{figure}
\figurenum{9}
\plotone{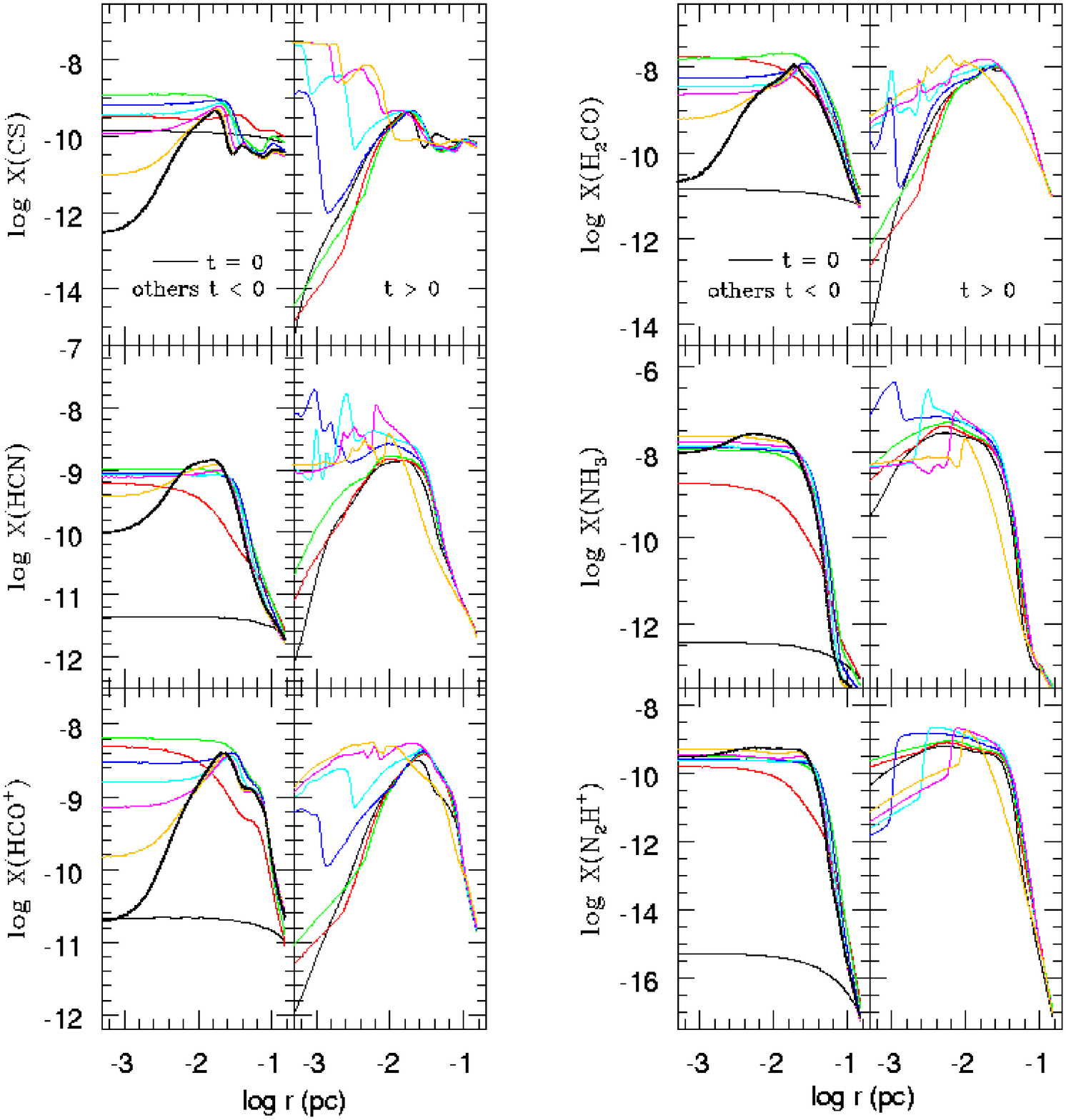}
\figcaption{ 
The evolution of abundance profiles. The left box for each 
molecule shows the molecular abundance profiles before collapse, and the right
boxes represent the abundance profiles after collapse begins. 
The selected time steps in left boxes are $5\times10^5$ (thin black), 
$2.5\times10^5$ (red), 
$1.25\times10^5$ (green), $6.5\times10^4$ (blue), $3.5\times10^4$ (cyan), 
$1.5\times10^4$ (magenta), $5\times10^3$ (orange) years before collapse, 
and the thick black line indicates the abundance profile at the 
initiation of collapse (t$=$0 year).  
Time steps in right box are $10^3$ (black), $10^4$ (red), $2\times10^4$ (green),
$5\times10^4$ (blue), $10^5$ (cyan), $2.0\times10^5$ (magenta), and 
$5\times10^6$ (orange) years after collapse.   
Moves for all time steps are found in 
{\bf http://peggysue.as.utexas.edu/sf/CHEM/}.
}
\end{figure}

\begin{figure}
\figurenum{10}
\plotone{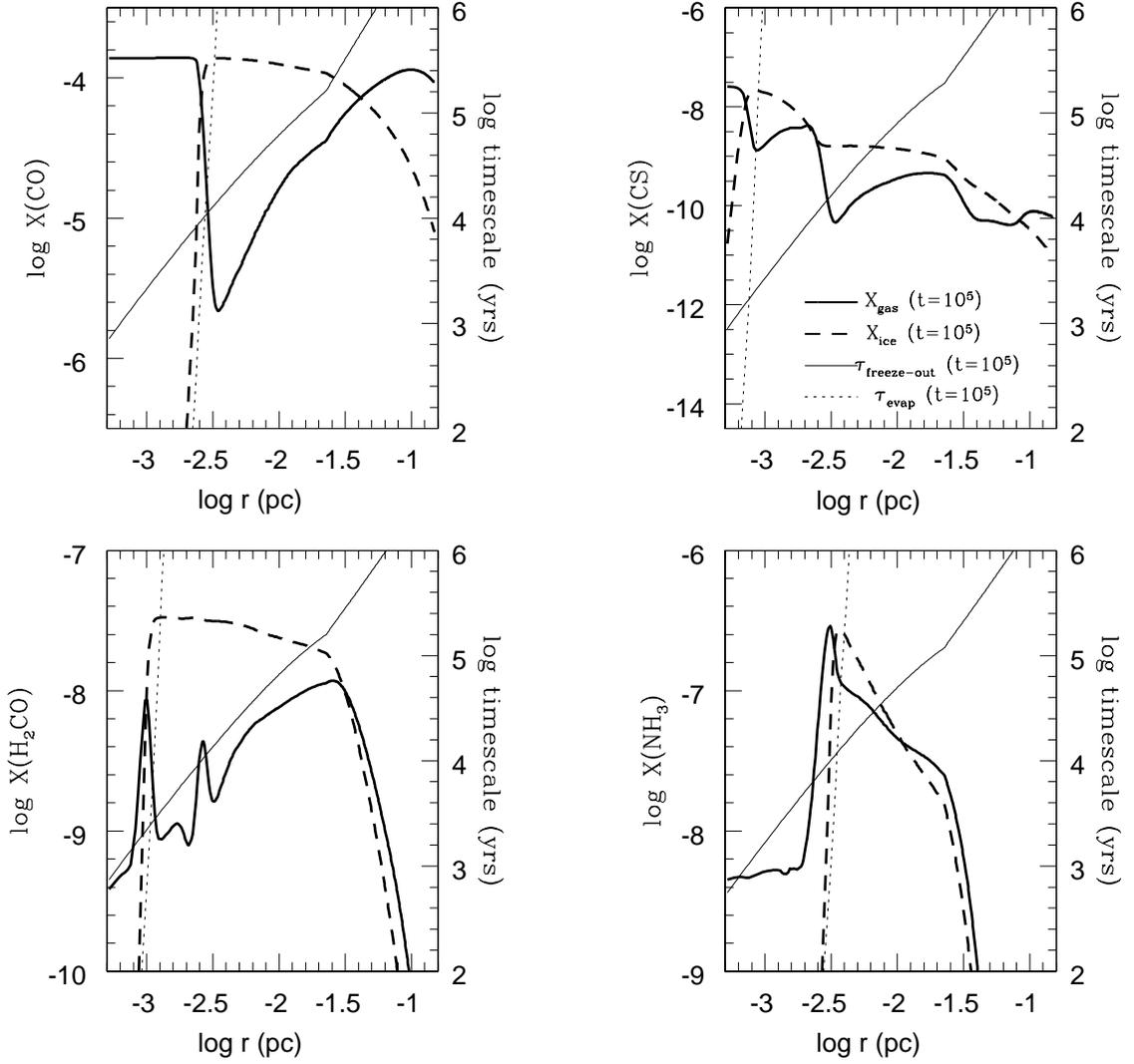}
\figcaption{The comparison of evaporation timescale and freeze-out timescale.
Thick solid and dashed lines represent abundance profiles in gas and ice at
$t=100,000$ years. 
Thin solid and dotted lines indicate the 
freeze-out timescale and evaporation timescale, respectively, at $t=100,000$ 
years.
At the point when two timescale are same, molecules are desorbed out from dust
grain surfaces quickly producing a sharp evaporation front. 
The evaporation front moves outward with time.
}
\end{figure}

\begin{figure}
\figurenum{11}
\plotone{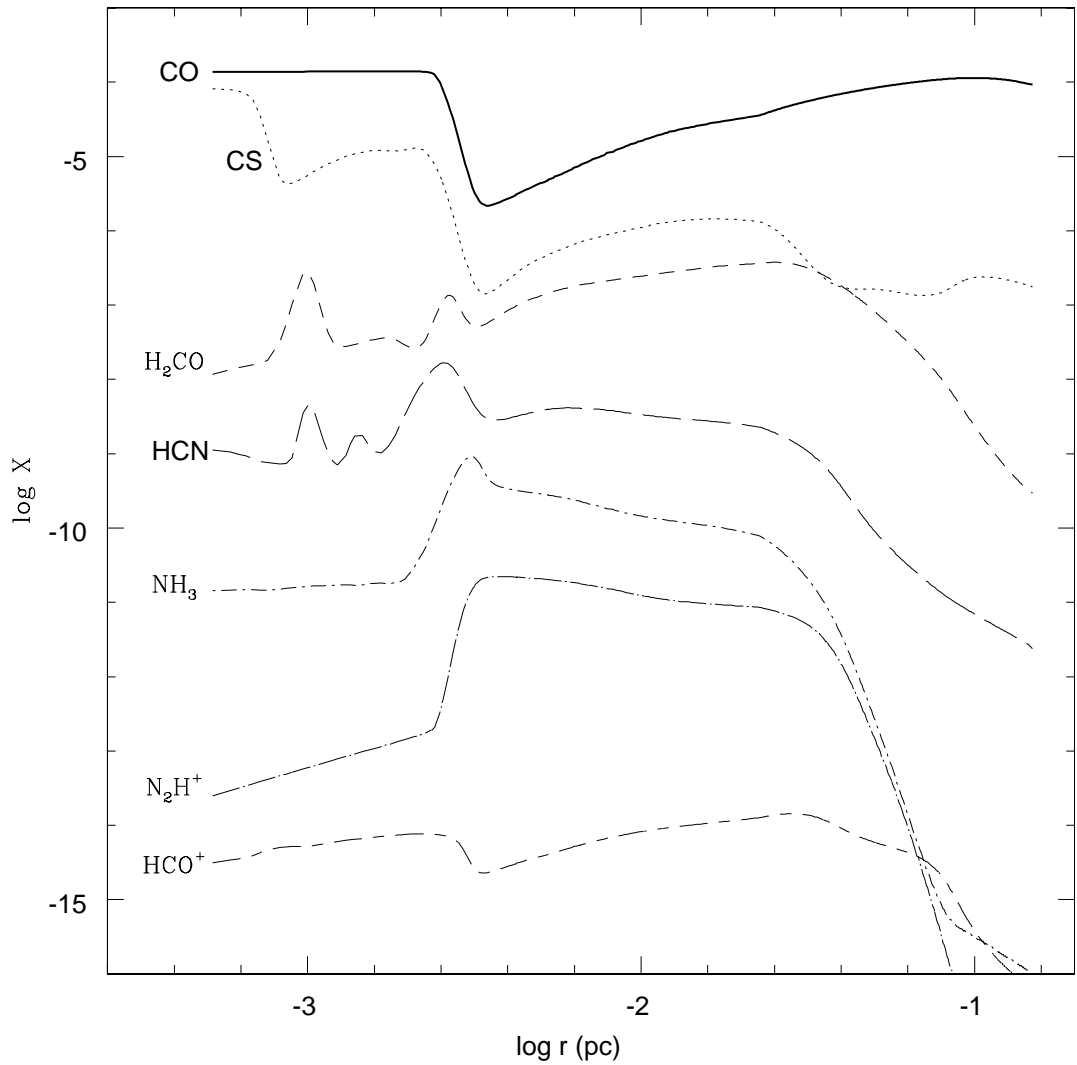}
\figcaption{Abundance profiles at a given time, $t=100,000$ years. The 
distributions of other molecules are mainly dependent on that of CO.  
The abundances decrease at the outer radii because molecules are easily 
photodissociated for low external extinction (0.5 mag).
Profiles have been shifted up and down to compare with the CO abundance profile
better, so y-axis does not give correct abundances of molecules except for
CO and HCN. 
CS and H$_2$CO are shifted up by the 3.5 and 1.5 orders of magnitude, 
and NH$_3$, $\rm N_2H^+$, and HCO$^+$ are shifted down by the 2.5, 2, and 5.5 
orders of magnitude, respectively.
}
\end{figure}

\begin{figure}
\figurenum{12}
\plotone{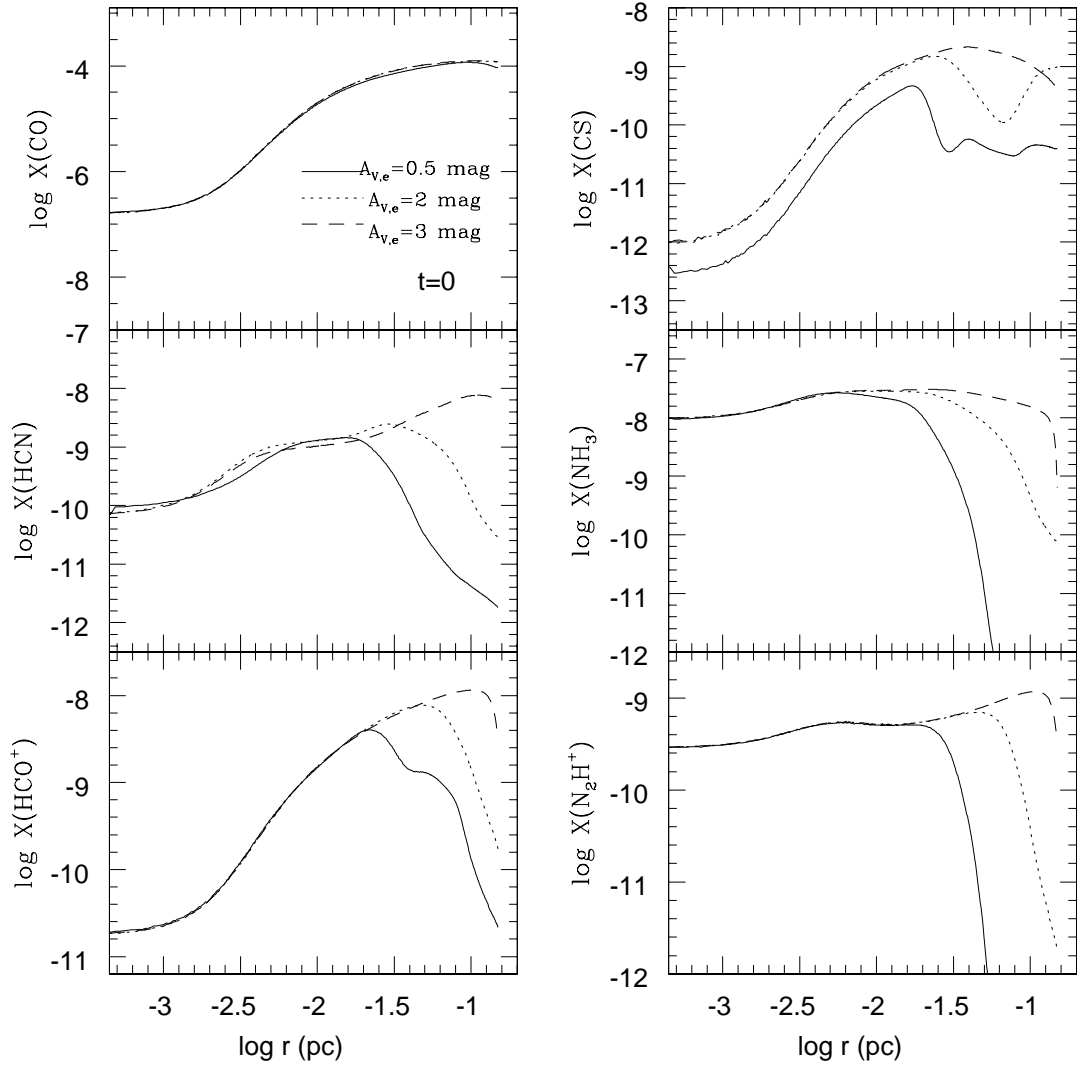}
\figcaption{Abundance profiles in models with different external visual 
extinctions, at the initiation of collapse. 
}
\end{figure}

\begin{figure}
\figurenum{13}
\plotone{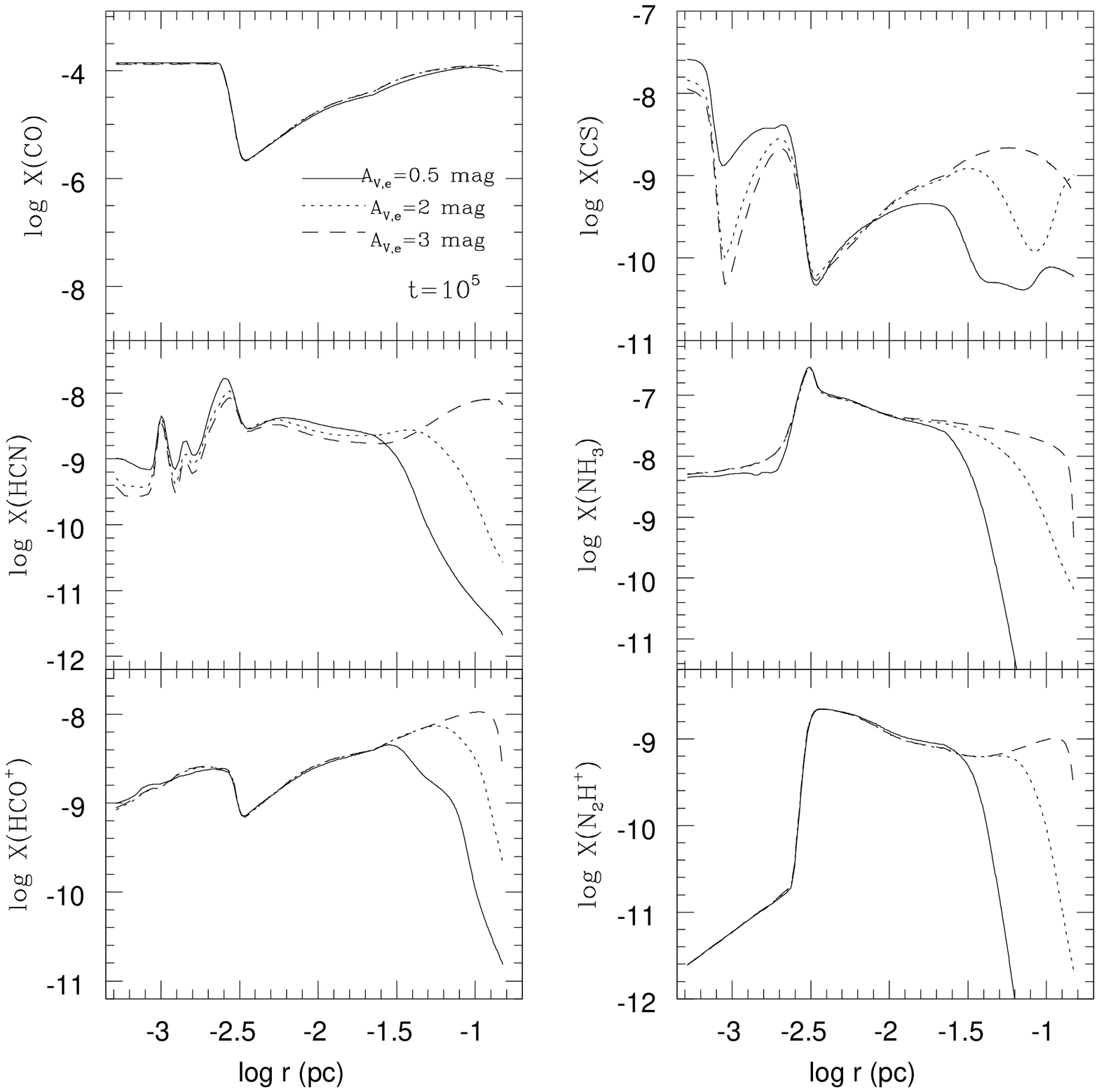}
\figcaption{ 
The same as in Figure 12, but at $t=100,000$ years. 
}
\end{figure}

\begin{figure}
\figurenum{14}
\plotone{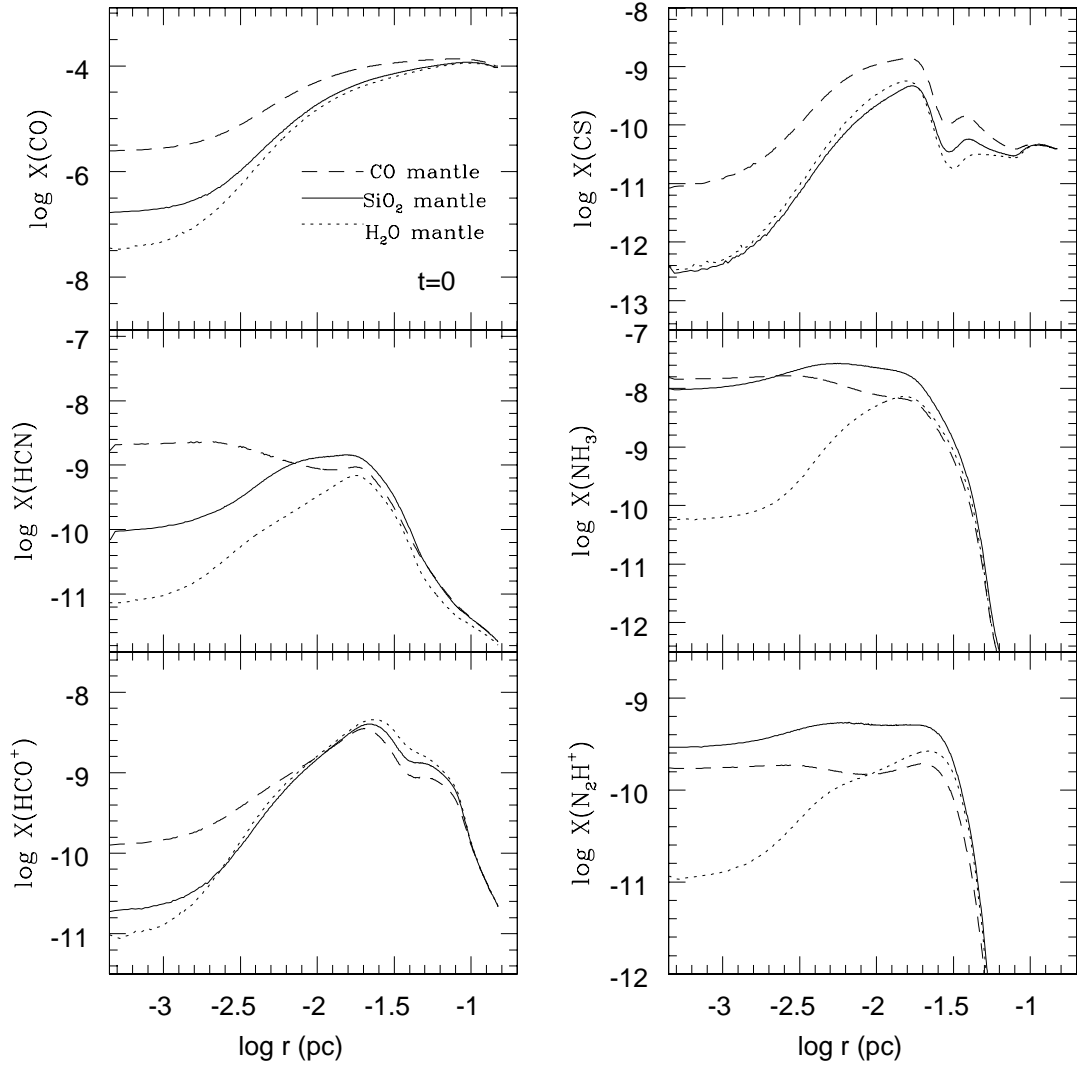}
\figcaption{Comparison between different binding energies onto dust grain
surfaces at the initiation of collapse.  
Solid line indicates the result of our fiducial model with the binding energy
onto SiO$_2$ grain surfaces. Dotted line represents the result of a model
with higher binding energy onto water dominant grain surfaces.
Dashed line is for lower binding energy onto CO dominant grain surfaces.
}
\end{figure}

\begin{figure}
\figurenum{15}
\plotone{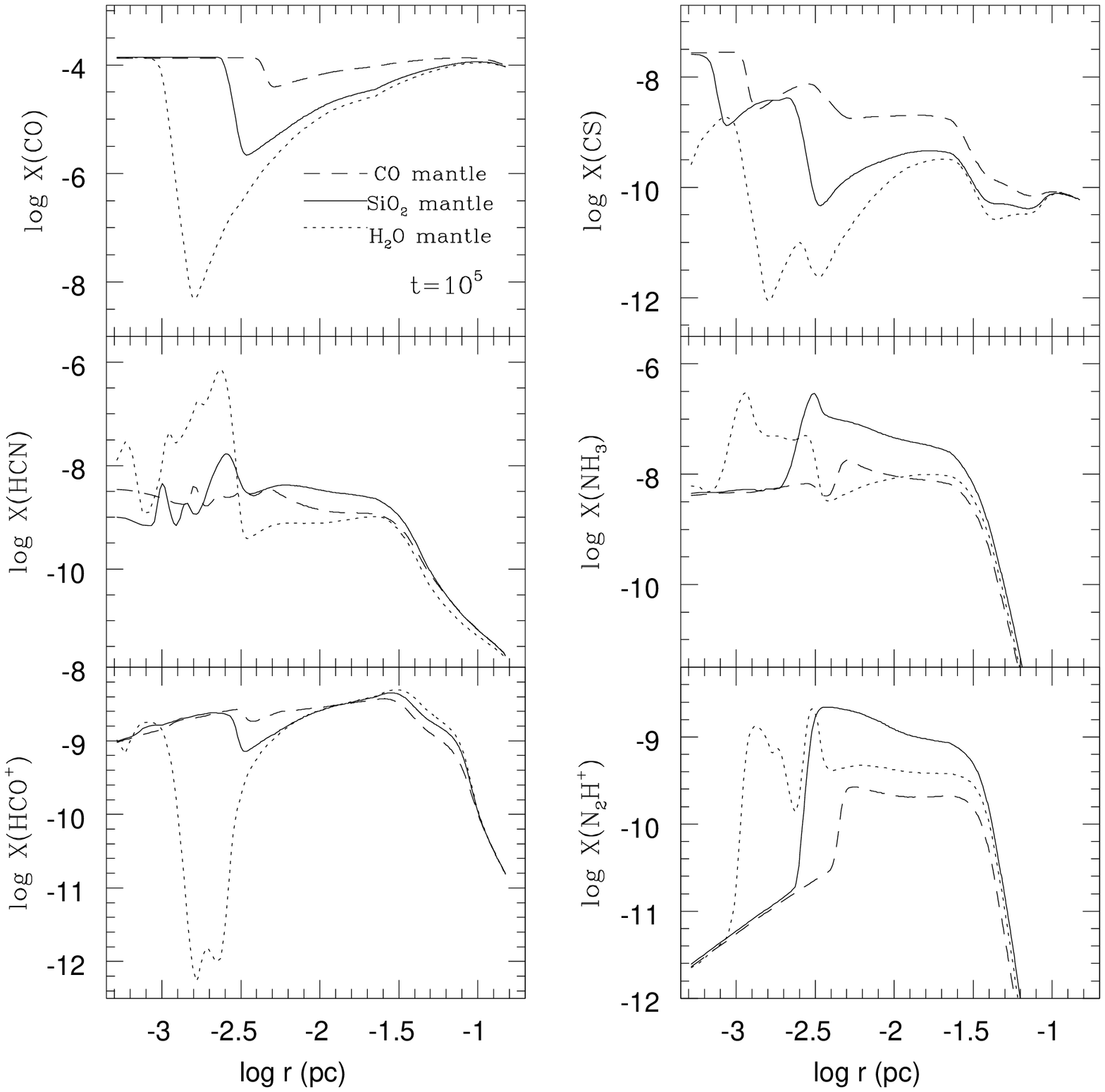}
\figcaption{The same as in Figure 14, but at $t=100,000$ years.
}
\end{figure}

\begin{figure}
\figurenum{16}
\plotone{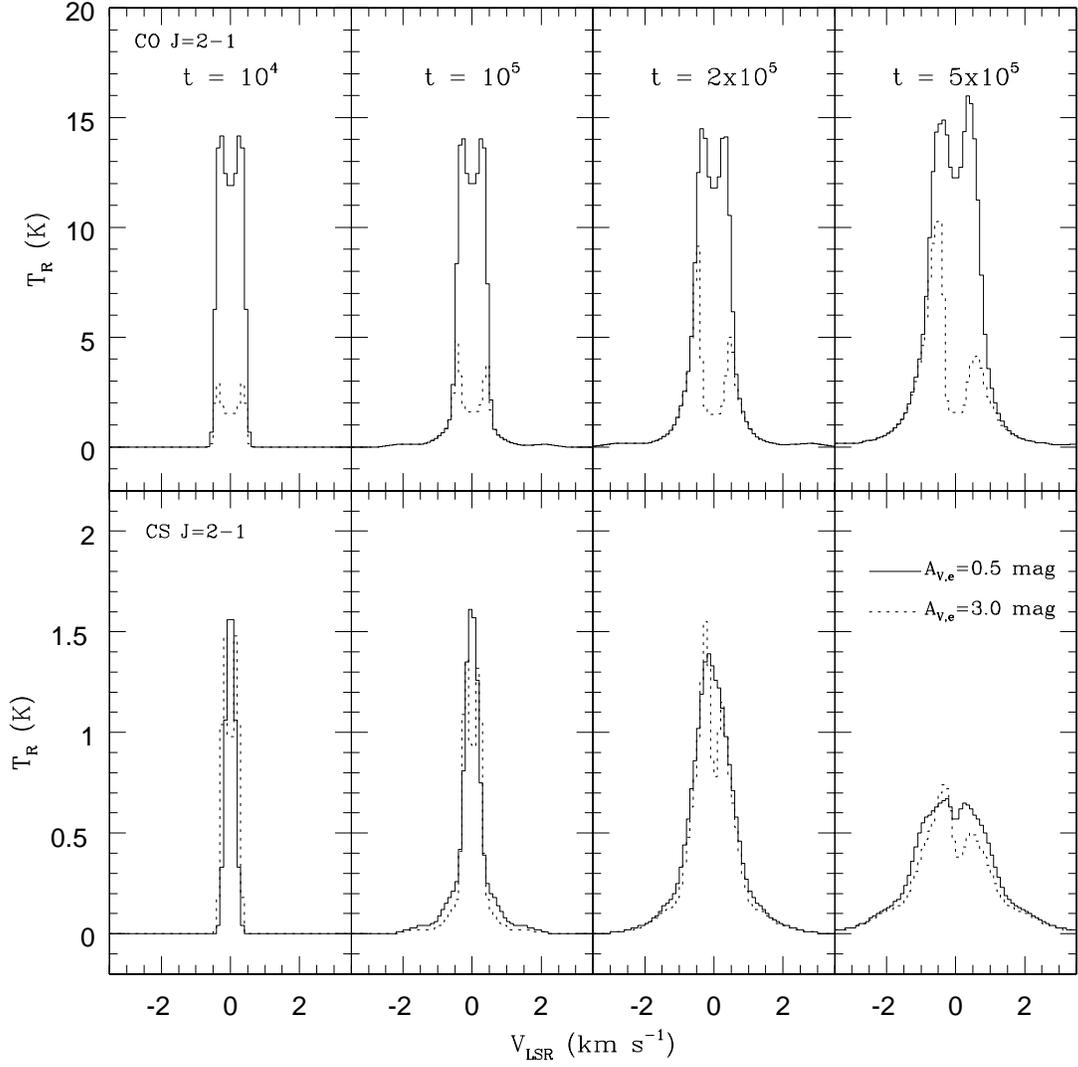}
\figcaption{The evolution of CO 2-1 and CS 2-1 lines. 
Solid and dotted lines represent line profiles simulated from the chemical 
model with 0.5 mag and 3 mag of external visual extinctions, respectively. 
Numbers in upper part of boxes indicate time steps. Line profiles at
the earlier time steps than 10,000 years are very similar to line profiles
at 10,000 years.
}
\end{figure}

\begin{figure}
\figurenum{17}
\plotone{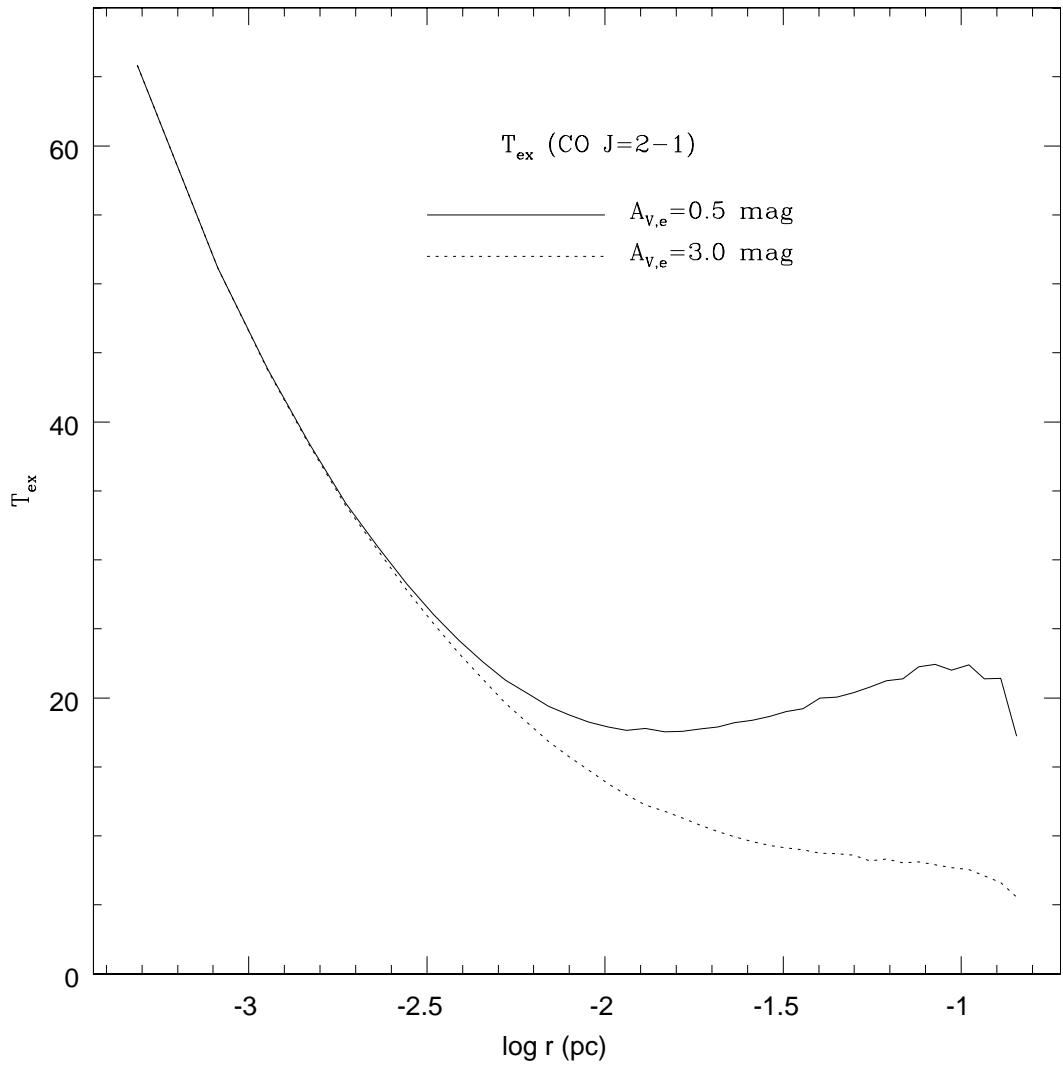}
\figcaption{ The profiles of excitation temperature of the CO 2$-$1 line 
in models with different external visual extinctions at 500,000 years.
}
\end{figure}

\clearpage

\begin{figure}
\figurenum{18}
\plotone{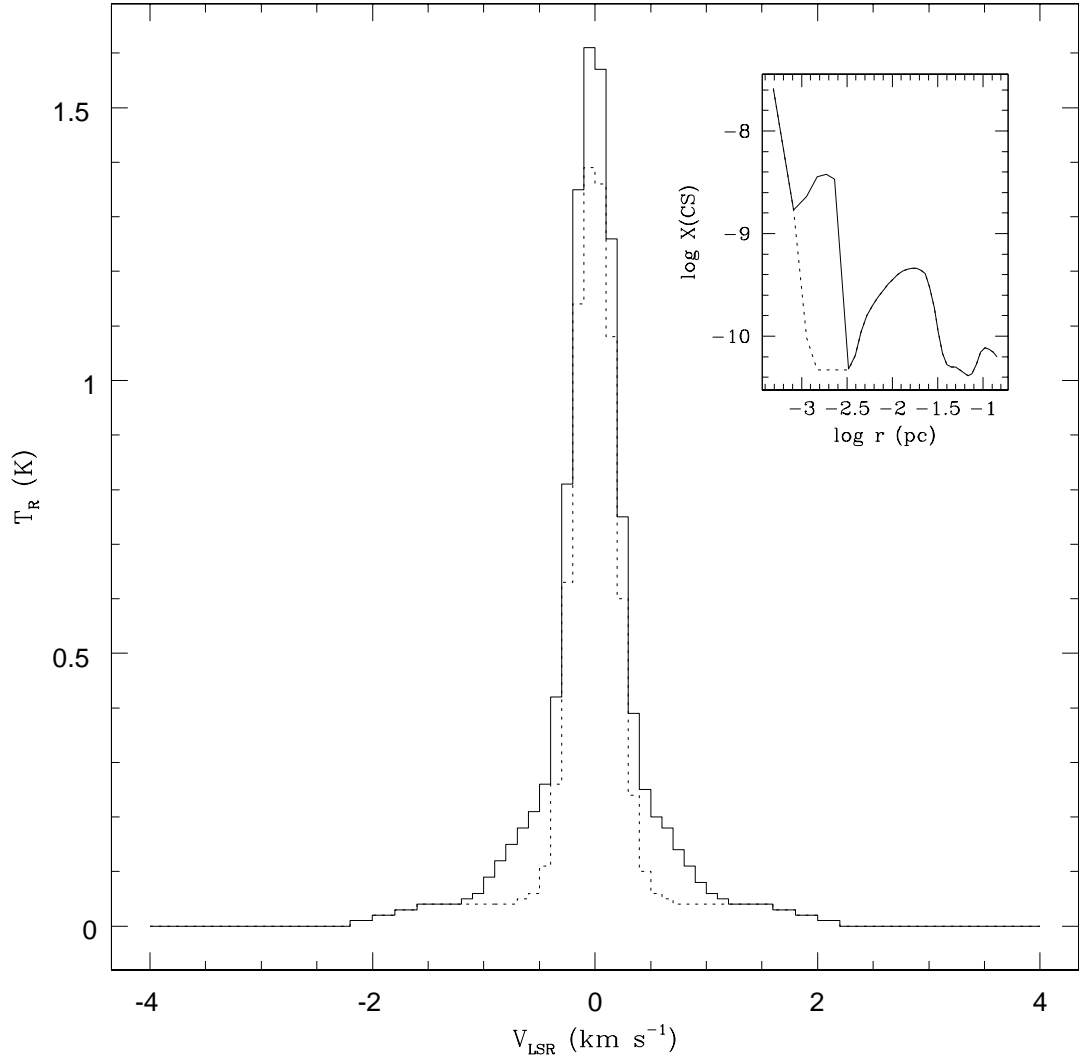}
\figcaption{
The comparison of the CS J$=2-1$ lines with different abundance profiles at
t$=$100,000 years.
Solid line has been simulated with the actual abundance profile obtained from
chemical model, and dotted line has been simulated with the abundance profile
where the second bump has been removed artificially. 
The solid line and dotted in the small box indicate the modeled abundance 
profile and the artificial abundance profile, respectively.   
}
\end{figure}

\begin{figure}
\figurenum{19}
\plotone{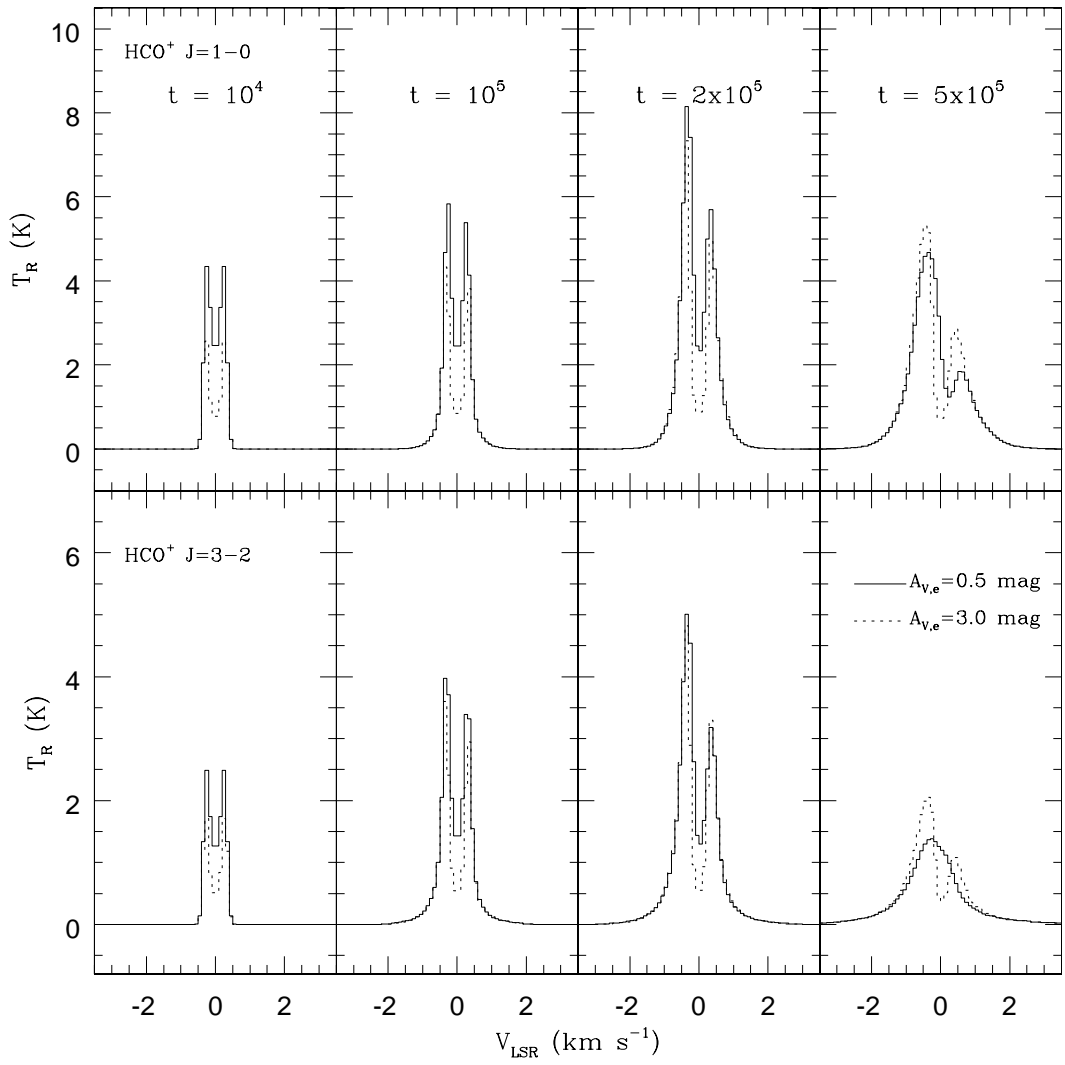}
\figcaption{The evolution of HCO$^+$ 1$-$0 and 3$-$2 lines. 
Solid and dotted lines represent line profiles simulated from the chemical
models with 0.5 mag and 3 mag of external visual extinctions, respectively.
}
\end{figure}

\begin{figure}
\figurenum{20}
\plotone{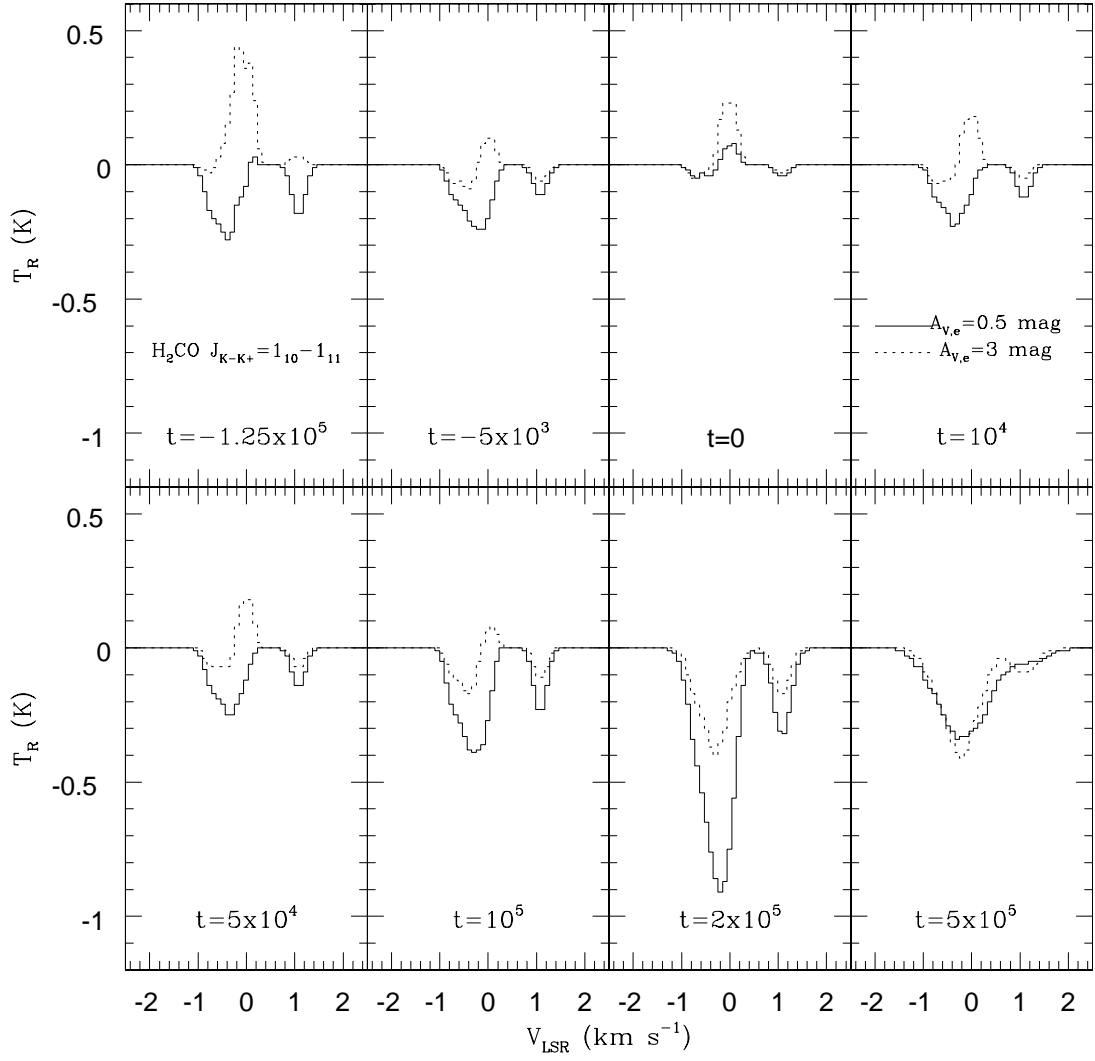}
\figcaption{  
The evolution of H$_2$CO 6 cm line.
Solid and dotted lines represent line profiles simulated with the results of 
chemical models of $\rm A_{V,e}=0.5$ and 3.0 mag, respectively.
}
\end{figure}

\clearpage

\begin{figure}
\figurenum{21}
\plotone{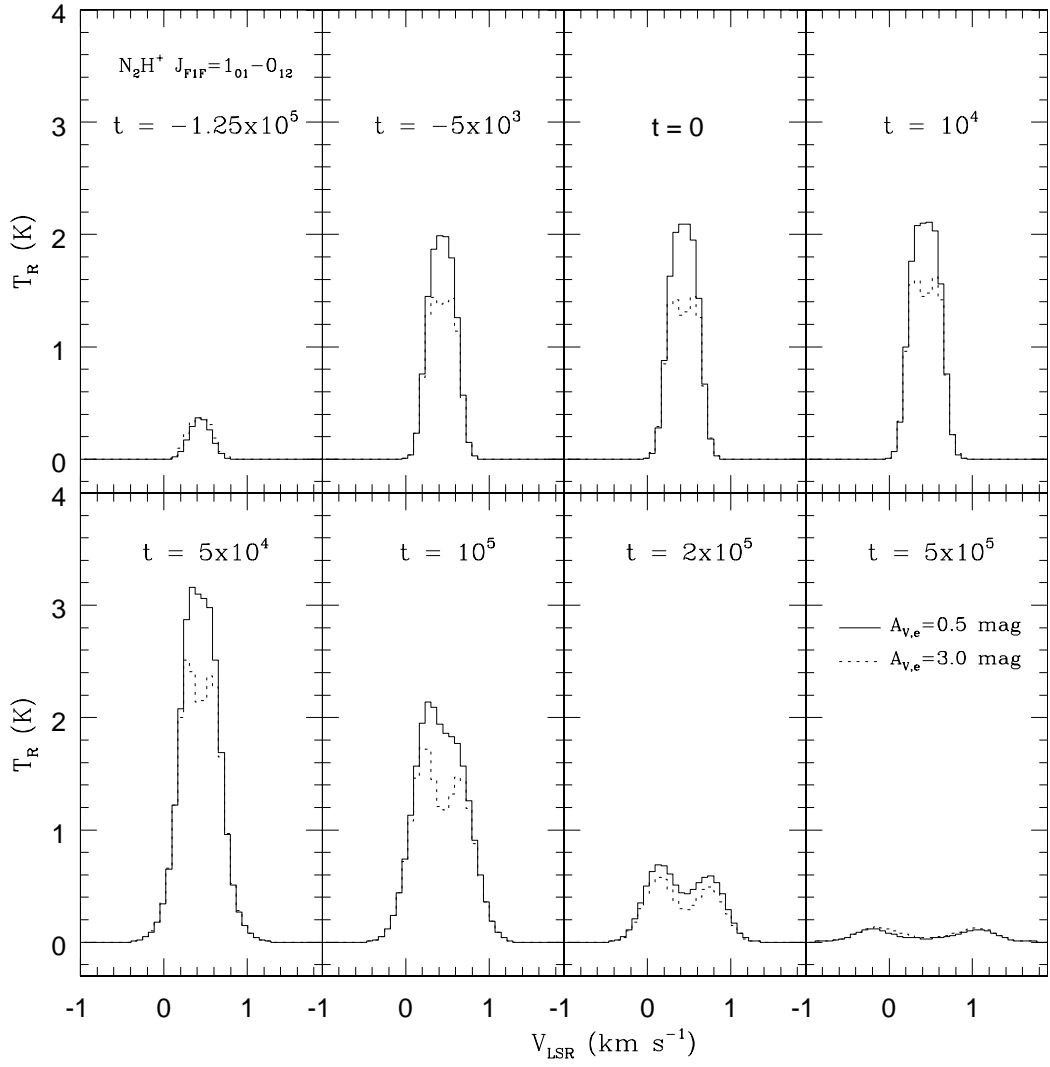}
\figcaption{ The evolution of $\rm N_2H^+$ isolated component ($1_{01}-0_{12}$).
}
\end{figure}

\begin{figure}
\figurenum{22}
\plotone{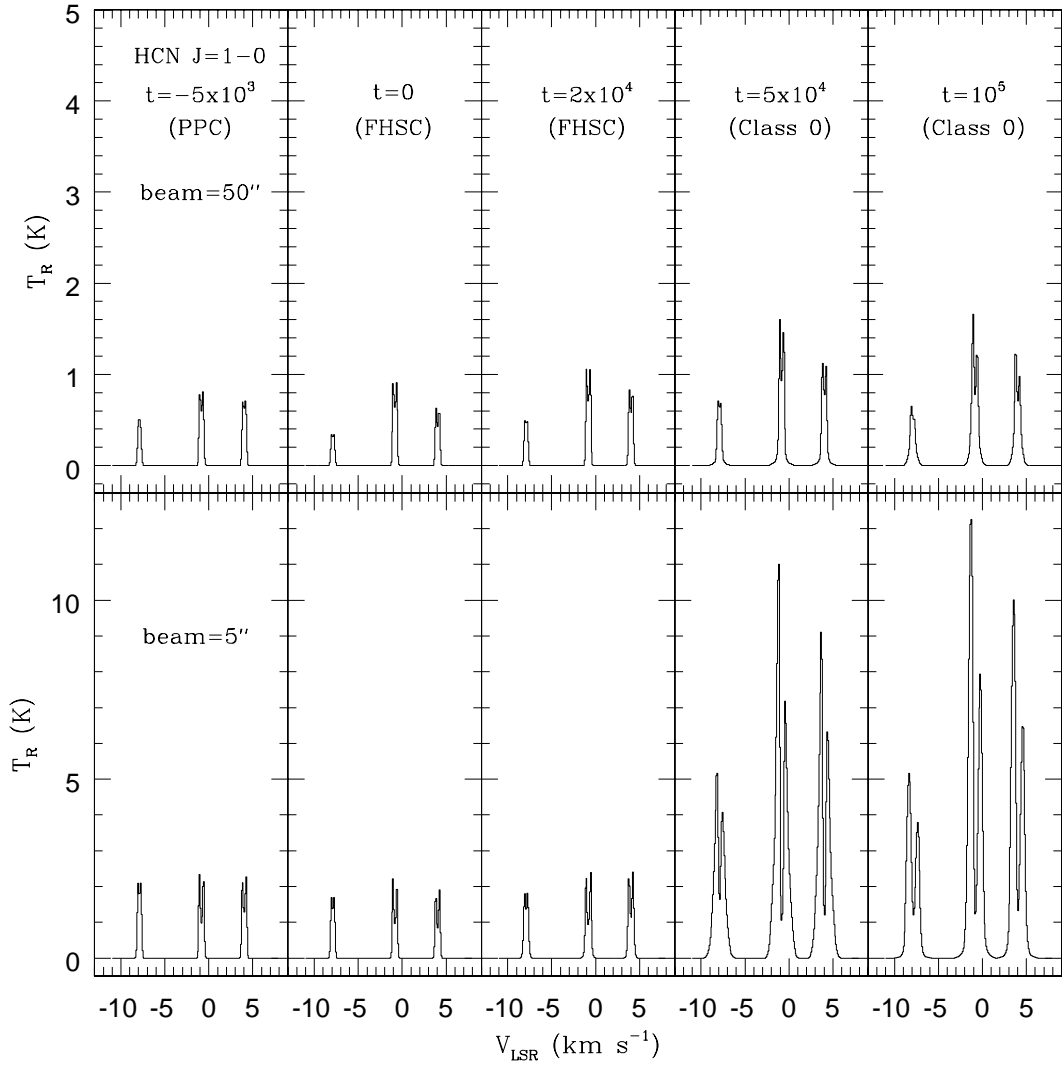}
\figcaption{ Comparison of HCN $1-0$ lines simulated with different
resolutions.
}
\end{figure}

\begin{figure}
\figurenum{23}
\plotone{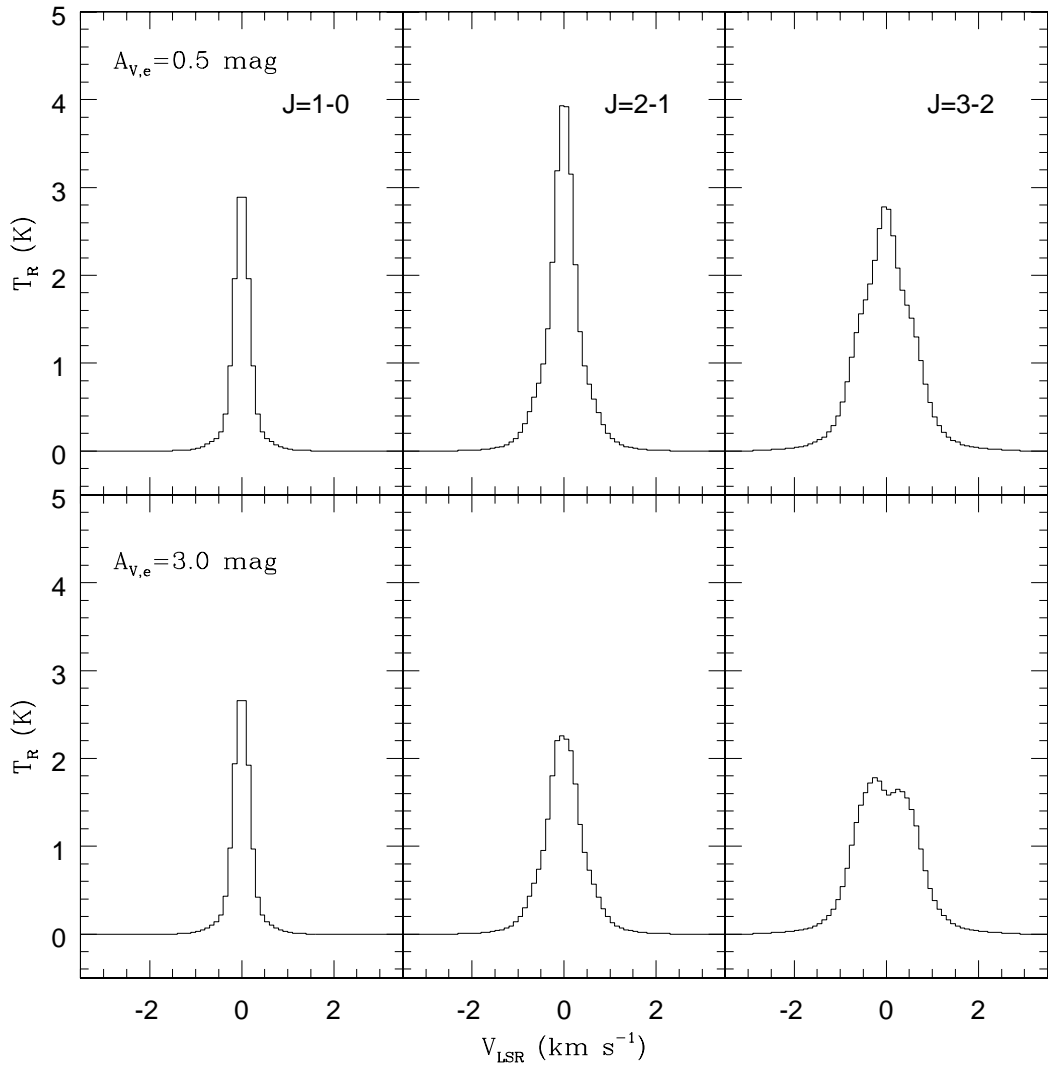}
\figcaption{ Comparison of C$^{18}$O $1-0$, $2-1$, and $3-2$ in 
$\rm A_{V,e}=0.5$ and 3 mag at $t=200,000$ years. 
}
\end{figure}

\begin{figure}
\figurenum{24}
\plotone{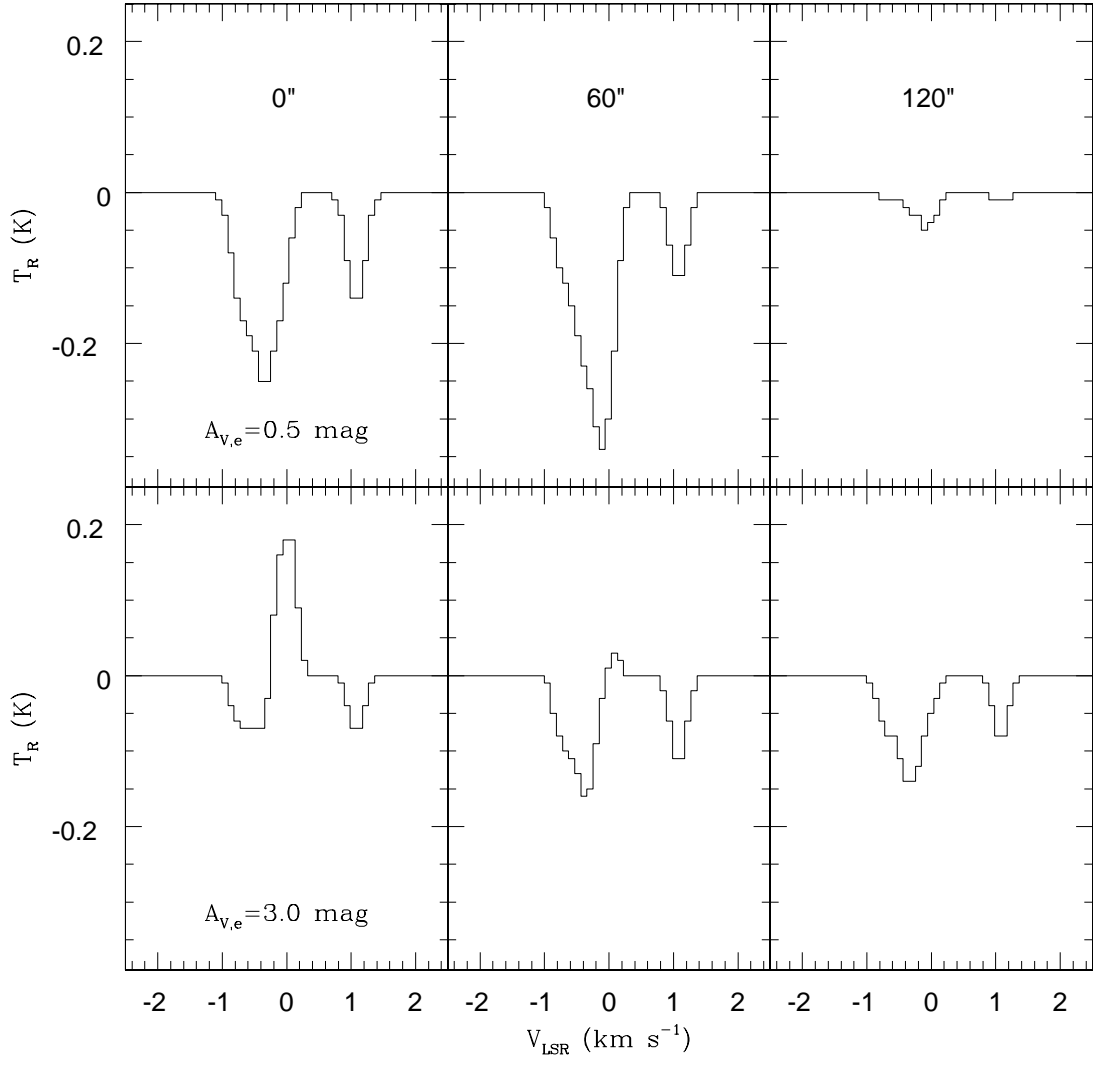}
\figcaption{ Comparison of ortho-H$_2$CO 6 cm line in $\rm A_{V,e}=0.5$ and 3 
mag at 50,000 years in the assumption of the same dust properties. 
}
\end{figure}

\begin{figure}
\figurenum{25}
\plotone{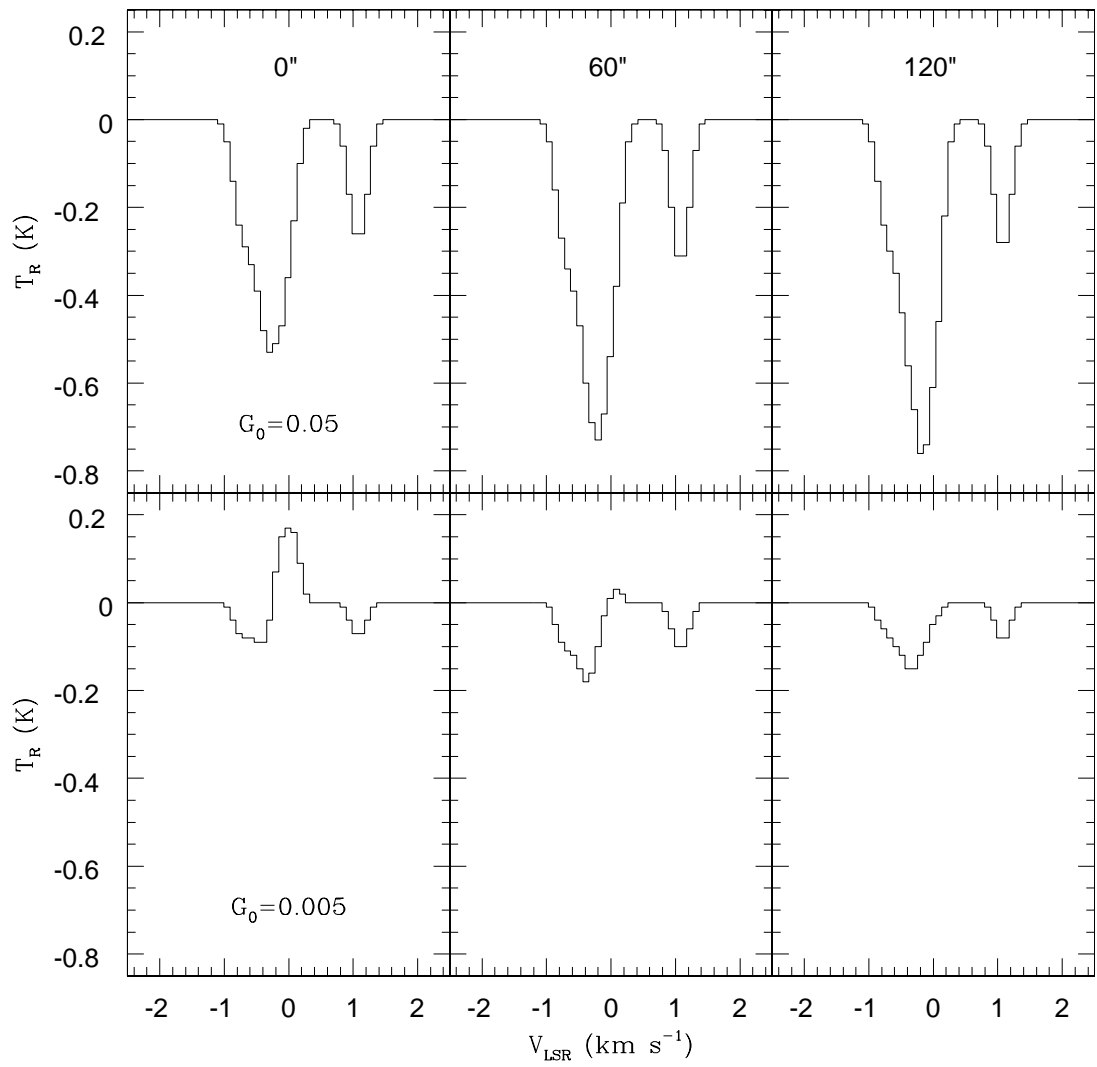}
\figcaption{ Comparison of ortho-H$_2$CO 6 cm line in $G_0=0.05$ and 0.005 
with the same external visual extinction of 3 mag at 50,000 years. 
}
\end{figure}

\begin{figure}
\figurenum{26}
\plotone{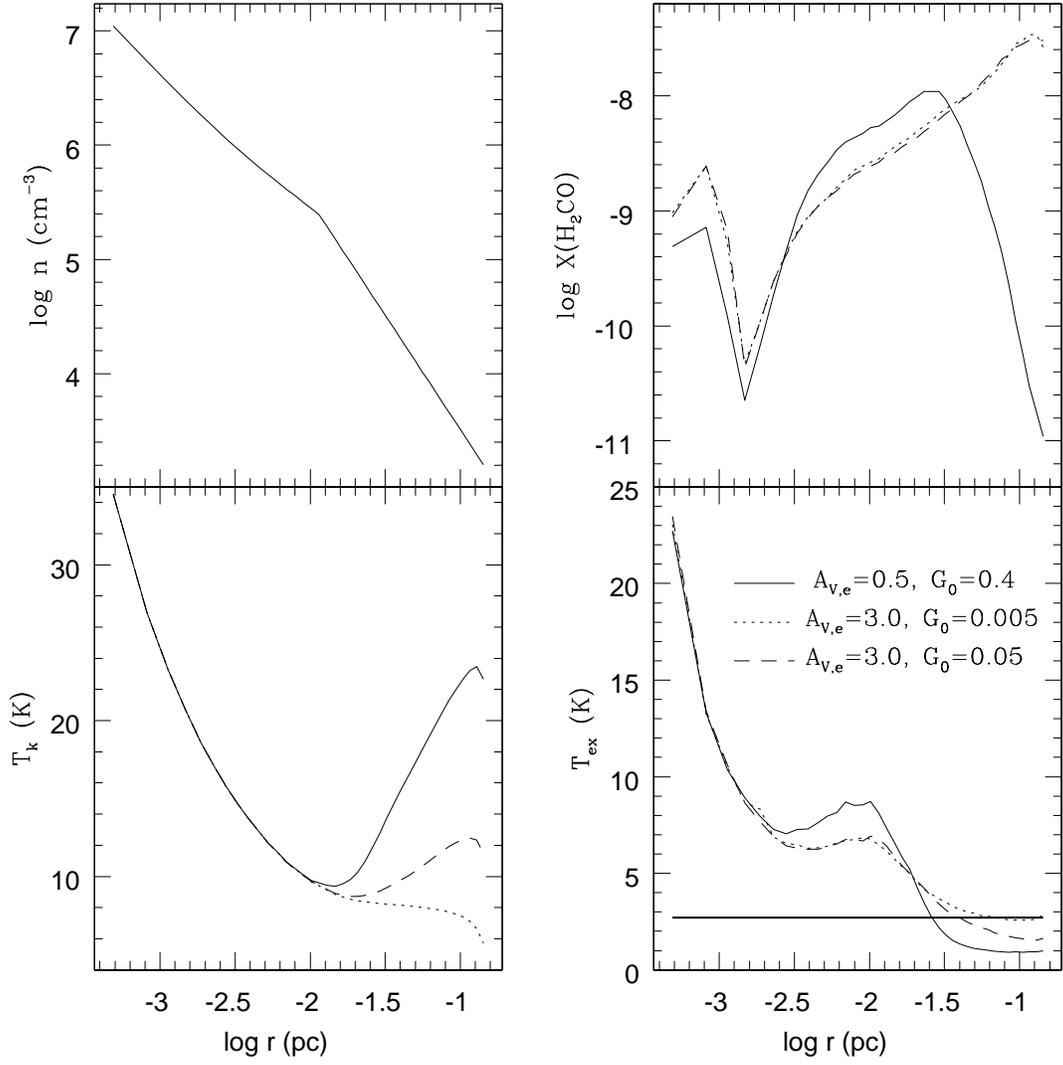}
\figcaption{Models of various $\rm A_{V,e}$ and $G_0$ at $t=50,000$ years. 
In the right bottom box, lines represent the excitation temperatures of the 
strongest component in the hyperfine structure of the ortho-H$_2$CO 6 cm line.
The thick horizontal line indicates the temperature of the CBR (2.7 K).
}
\end{figure}

\end{document}